%% file: main.tex
\documentclass[usenatbib]{mnras}

\usepackage{graphicx}
\usepackage{cleveref}
\usepackage{url}
\usepackage{amsmath}
\usepackage[T1]{fontenc}
\usepackage{aecompl}
\usepackage{color}

\graphicspath{{fig/}}
\newcommand{\WISE}{\mbox{\textit{WISE}}}
\newcommand{\GALEX}{\mbox{\textit{GALEX}}}
\newcommand{\ATLAS}{\mbox{ATLAS\textsuperscript{3D}}}
\newcommand{\Rampazzo}{\citet{Rampazzo13}}
\newcommand{\Chabrier}{\citet{Chabrier03}}
\newcommand{\Salpeter}{\citet{Salpeter55}}
\newcommand{\Marigo}{\citet{Marigo08}}

\newcommand{\ISO}{\mbox{\textit{ISO}}}
\newcommand{\TM}{\mbox{2MASS}}
\newcommand{\Spitzer}{\mbox{\textit{Spitzer}}}
\newcommand{\Herschel}{\mbox{\textit{Herschel}}}
\newcommand{\IRS}{\mbox{\Spitzer-IRS}}

\newcommand{\SAURON}{\mbox{SAURON}}
\newcommand{\Class}[1]{Class-#1}
\newcommand{\Ks}{K\textsubscript{s}}
\newcommand{\IRAF}[1]{\textit{#1}}
\newcommand{\ellipse}{\IRAF{ellipse}}
\newcommand{\fitsky}{\IRAF{fitsky}}
\newcommand{\kms}{km s\(^{-1}\)}
\newcommand{\KT}{K20}

\newcommand{\lineratio}[2]{\mbox{#1/#2}}

\newcommand{\sixtwo}{\mbox{6.2 \micron}}
\newcommand{\sevenseven}{\mbox{7.7 \micron}}
\newcommand{\eightsix}{\mbox{8.6 \micron}}
\newcommand{\eleventhree}{\mbox{11.3 \micron}}
\newcommand{\twelveseven}{\mbox{12.7 \micron}}
\newcommand{\seventeen}{\mbox{17 \micron}}
\newcommand{\ratiosixseven}{\lineratio{\sixtwo}{\sevenseven}}
\newcommand{\ratioseveneleven}{\lineratio{\sevenseven}{\eleventhree}}
\newcommand{\ratioeightseven}{\lineratio{\eightsix}{\sevenseven}}
\newcommand{\ratioelevenseven}{\lineratio{\eleventhree}{\sevenseven}}
\newcommand{\ratiotwelveeleven}{\lineratio{\twelveseven}{\eleventhree}}

\newcommand{\ratioseventeeneleven}{\lineratio{\seventeen}{\eleventhree}}

\newcommand{\htwo}{\mbox{H\textsubscript{2}}}
\newcommand{\eighteenseven}{\mbox{[\ion{S}{iii}] 18.7 \micron}}
\newcommand{\thirtythreefive}{\mbox{[\ion{S}{iii}] 33.5 \micron}}
\newcommand{\molecularthree}{\mbox{\htwo S(3)}}
\newcommand{\molecularone}{\mbox{\htwo S(1)}}

\newcommand{\ratiosulfur}{\lineratio{\eighteenseven}{\thirtythreefive}}
\newcommand{\ratiomolecular}{\lineratio{\molecularthree}{\molecularone}}

\crefname{figure}{Fig.}{Figs.}
\Crefname{figure}{Fig.}{Figs.}
\crefname{table}{Table}{Tables}
\Crefname{table}{Table}{Tables}
\crefname{section}{Section}{Sections}
\Crefname{section}{Section}{Sections}

\defcitealias{Planck14}{Planck Collaboration 2014}
\defcitealias{Astropy}{Astropy Collaboration 2013}

\title[ETGs in \GALEX{} and \WISE{}]{Circumstellar Dust, PAHs, and
  Stellar Populations in \\
  Early-Type Galaxies: Insights from \GALEX{} and \WISE{}}
\author[Simonian \& Martini]{Gregory V. Simonian\(^{1}\)\thanks{E-mail:
simonian@astronomy.ohio-state.edu} and Paul Martini\(^{1,2}\)
    \\
    \(^{1}\)Department of Astronomy, The Ohio State University, 140 West 
    18th Avenue, Columbus, OH 43210, USA\\
    \(^{2}\)Center for 
    Cosmology and AstroParticle Physics (CCAPP), The Ohio State 
    University, 191 W.  Woodruff Ave., Columbus, OH 43210, USA}

\begin{document}

\maketitle

\begin{abstract}
A majority of early-type galaxies contain interstellar dust, yet the
origin of this dust, and why the dust sometimes exhibits unusual
polycyclic aromatic hydrocarbon (PAH)
ratios, remains a mystery. If the dust is internally produced, it likely
originates from the large number of asymptotic giant branch stars associated 
with the old
stellar population. We present \GALEX{} and \WISE{} elliptical aperture
photometry of $\sim310$ early-type galaxies with Spitzer mid-infrared
spectroscopy and/or ancillary data from \ATLAS{}, to characterize their
circumstellar dust and the shape of the radiation field that illuminates
the interstellar PAHs. We find that circumstellar dust is ubiquitous in
early-type galaxies, which indicates some tension between stellar
population age estimates and models for circumstellar dust production in
very old stellar populations. We also use dynamical masses from
\ATLAS{} to show that \WISE{} W1 (3.4 $\mu$m) mass-to-light ratios are 
consistent with the initial mass function variation found by previous work.
While the stellar population differences in early-type galaxies
correspond to a range of radiation field shapes incident upon the
diffuse dust, the ratio of the ionization-sensitive $7.7\mu$m to
$11.3\mu$m PAH feature does not correlate with the shape of the
radiation field, nor to variations with the size-sensitive $11.3\mu$m to
$17\mu$m ratio. The $7.7\mu$m to $11.3\mu$m PAH ratio does tend to be
smaller in galaxies with proportionally greater H$_2$ emission, which is
evidence that processing of primarily smaller grains by shocks is
responsible for the unusual ratios, rather than substantial differences
in the overall PAH size or ionization distribution.
\end{abstract}


\section{Introduction}

While early-type galaxies (ETGs) were historically associated with
uniformly old stellar populations and no cold interstellar medium
(ISM), forty years of multiwavelength observations have demonstrated that 
view is too simplistic. Instead, many ETGs have a complex, multi-phase ISM, 
often with a mixture of cold \citep{Knapp85,Wardle86}, warm
\citep{Caldwell84,Phillips86,Sadler87}, and hot
\citep{Forman85,Canizares87} gas. 
Improved angular resolution has also led to the detection of dust lanes in 
many ETGs \citep{Sadler85,Sparks85,Ebneter88,Veron88,Goudfrooij94b}, and
observations in the far-infrared (FIR) indicated
that many ETGs contain a diffuse, cold dust component
\citep{Jura87,Knapp89,Goudfrooij95,Bregman98}.  
Mid-Infrared (MIR) observations from the \textit{Infrared Space
Observatory} (\ISO{}) satellite found flux in excess 
of expectations for the stellar population of
many ETGs, which was postulated to arise from either Polycyclic Aromatic
Hydrocarbon (PAH) or Very Small Grain (VSG)
emission \citep{Madden99,Ferrari02,Xilouris04,Pahre04}.
The existence of these small grains was surprising because of
their short lifetimes in hot plasma \citep{Draine79,Dwek92}, and so the
origin of these grains is hotly debated.

MIR spectroscopy of individual PAH features in ETGs indicated that the 
relative strengths of short-wavelength and long-wavelength PAH
features were often vastly reduced compared to the same features in 
star-forming galaxies
\citep{Kaneda05}. Proposed physical conditions which could lead to the 
relatively weaker short-wavelength features include a grain population 
dominated by neutral rather than ionized PAHs, and a larger grain size
distribution compared to star-forming galaxies \citep{Draine07}. 

An explanation for both the existence of PAHs and their anomalous line ratios 
in ETGs has proven elusive. \citet{Kaneda08} found that PAH emission 
is uncorrelated with stellar emission, and suggested that
larger neutral PAHs were externally accreted through mergers.
Alternatively, \citet{Vega10} posited that PAHs
are produced by carbon stars formed in a minor star-forming event, and
that shocks with the ambient ISM then preferentially destroyed the small grains.

For galaxies without interstellar dust, we get a direct view of the MIR
emission of the stellar population, which makes these galaxies well-suited to 
test Stellar Population Synthesis (SPS) models. They are representative of old 
stellar populations with little ongoing star formation 
\citep{Yi05,Kaviraj07,Temi09,Shapiro10}, and negligible extinction due
to dust. We will use them to test stellar models for the evolved giant stars that dominate their light,
especially the substantial progress on the
Thermally Pulsating Asymptotic Giant Branch (TP-AGB) phase over the last decade 
\citep{Marigo08,Girardi10,Cassara13,Marigo13,Rosenfield14}. This phase is 
critical as the stars can contribute a significant amount of the integrated 
flux in the infrared 
\citep{Maraston05, Kelson10, Melbourne12, Conroy13, Melbourne13}.

Previous studies of AGB stars have shown that circumstellar dust is
necessary to adequately describe their spectra in the MIR 
\citep{Bedijn87,vanLoon99,Trams99}. However, the inclusion of
circumstellar dust within SPS models has been difficult, and only a few
models incorporate dusty AGB stars into their spectral libraries
\citep{Bressan98,Silva98,Marigo08,Villaume15}. The expansion of MIR
observations provide new opportunities to compare these circumstellar dust 
models to real stellar populations \citep[e.g.][]{Norris14,Villaume15}.

It is difficult to compare circumstellar dust models to data in active 
star-forming galaxies because the circumstellar dust emission is often
dwarfed by emission from dust in the diffuse ISM\@.
For our sample of passive early-type galaxies with much less interstellar 
dust, the circumstellar component can be observed in the MIR 
(beyond about 10 \micron{}). Excess flux associated with
circumstellar dust has been identified in a number of studies of
galaxies without evidence for interstellar dust 
\citep{Bressan98, Athey02, Martini13}. 

The MIR region from 3--5 \micron{} dominated by photospheric emission is 
valuable for stellar mass measurements. Since low-mass stars contain most of 
the stellar mass of a galaxy, observations at these wavelengths are more robust 
to variations in metallicity, star formation history (SFH), and star formation 
rate \citep[SFR;][]{Meidt14}.  Results from SPS models have traditionally been the only way 
to measure the stellar masses of large numbers of galaxies. Alternative mass 
estimates were recently released by \ATLAS{} for a volume-limited sample of 260 
ETGs closer than 42 Mpc with \(M_* \ga 6 \times 10^9 M_{\sun}\). 
\citet{Cappellari13a} derived dynamical masses for these galaxies using 
\textit{r}-band photometry, \SAURON{} integral-field unit (IFU) spectroscopy, 
and dynamical models based on the Jeans equations. They then obtained
stellar masses by subtracting a Navarro-Frenk-White (NFW) halo; which yielded 
stellar masses with assumptions independent from SPS models. 

We use MIR data from the \WISE{} satellite, which observed
the entire sky in four MIR bands: W1 (3.4 \micron), W2 (4.6 \micron), W3 (12
\micron), and W4 (22 \micron) (see \citet{Wright10} for further
details). The first two bands are similar to the 
\Spitzer{} Infrared Array Camera (IRAC) [3.6] and [4.5] bands, and the W4 band 
is similar to the Multiband Imaging Photometer on \Spitzer{} (MIPS) 24 
\micron{} band. In typical galaxies, the W1 and W2 bands are 
expected to trace the evolved stellar population; the W3 band 
will contain significant PAH features; and the W4 band will be dominated
by continuum emission from hot dust grains \citep[e.g.][]{Jarrett13}. Due to the 
all-sky coverage of \WISE{}, all sufficiently bright ETGs can be studied 
in the MIR, a substantial increase over previous targeted surveys. This wide coverage will yield 
valuable demographic data about the stellar populations, circumstellar
dust, and PAHs in ETGs, as well as identify promising targets for future
study with targeted missions.

We also include UV photometry from the \GALEX{} satellite to measure the shape
of the radiation field incident on any PAHs that may be present.
\GALEX{} observed 63\% of the sky to a depth of at least 
\(m_{AB}=20\) mag in the FUV (\(1516\mbox{\AA}\)) and NUV 
(\(2267\mbox{\AA}\)) bands, with a resolution
of about 4.25\arcsec{} and 5.25\arcsec{} respectively (see \citet{Morrissey05} and
\citet{Martin05} for further details). 

In this paper, we distinguish between ``interstellar'' and
``circumstellar'' dust as follows: ``circumstellar'' dust resides within
the stellar winds of AGB stars while ``interstellar'' dust resides within the 
diffuse ISM\@. Circumstellar dust is 
intrinsically connected to the stellar population, which makes it a valuable 
extension to Stellar Population Synthesis models. Meanwhile, interstellar 
dust is often uncorrelated with the stellar population and often
dominates the IR emission when it is present. 

Our paper is organized as follows: the next section of this paper contains a 
description of our samples of ETGs. \Cref{sec:processing} describes how we 
performed aperture photometry on both \WISE{} and \GALEX{}
images and how we distinguish between galaxies with and without diffuse
dust. In \Cref{sec:emission}, we calculate stellar mass-to-light ratios
for W1 from \ATLAS{} dynamical masses and compare them to mass-to-light
ratios predicted by SPS models. We
also compare models of circumstellar dust to our data, and use these to jointly 
constrain stellar ages and the masses of stars that produce
circumstellar dust, as well as investigate the extent to which circumstellar 
dust can contaminate measurements of the SFR as measured 
from MIR indicators. \Cref{sec:pahs} combines \GALEX{} and \WISE{} photometry 
with \Spitzer{} Infrared Spectrograph (\IRS{}) spectroscopy from previous 
works to investigate the properties of PAHs and their environments in ETGs. We 
also comment on the use of \WISE{} photometry to determine the MIR properties 
of ETGs.  We summarize our results in \cref{sec:conclusion}.

\section{Samples}
\label{sec:samples}

We use MIR and UV photometry to study circumstellar and interstellar
dust in ETGs. Our sample is drawn from two recent, comprehensive studies
of ETGs: the characterizations of interstellar dust in
ETGs with \IRS{} spectra \citep{Rampazzo13}, and the stellar
population and dynamical study from the \ATLAS{} survey.

\subsection{RSA \IRS{} Atlas}
\label{sec:rampazzo}

The Revised Shapley-Ames (RSA) catalog is a canonical collection of
bright, well-studied, nearby galaxies. \Rampazzo{} constructed their RSA
\IRS{} atlas by cross-matching the E--S0 galaxies from the RSA catalog 
with \IRS{} observations available in the \Spitzer{} Heritage Archive 
(SHA). This sample consists of 91 ETGs, including 56 E-type, 27 S0-type, and 
8 mixed E/S0+S0/E-type galaxies; their properties are given in \cref{tab:params}. 
\Rampazzo{} uniformly reprocessed and analyzed all of these spectra and 
measured line intensities for each of the detected emission lines. 

\Rampazzo{} classified galaxies according to the \citet{Panuzzo11}
classification scheme for MIR spectra. A summary of the classification scheme is as follows: 
\Class{0} galaxies are completely passive, that is apart from a few
broad circumstellar dust features \citep{Bressan98}, they have no
emission lines in the MIR\@. These galaxies have spectra consistent with
only an old 
stellar population. \Class{1} galaxies have emission features in their 
spectra, but no PAH features. \Class{2} and \Class{3} galaxies are those with
anomalous and normal PAH features, respectively; they will be the
primary focus of this work. Finally, \Class{4} galaxies have a hot
dust continuum. 

\subsection{The \ATLAS{} Sample}

The \ATLAS{} project surveyed 
a volume-limited sample within 42 Mpc of morphologically-selected 
early-type galaxies with 
\(M_K < -21.5\) mag \citep{Cappellari11}. This sample contains 260
ETGs: 68 E galaxies and 192 S0 galaxies (see \cref{tab:params}).
The extensive data available for the \ATLAS{} sample includes
Sloan Digital Sky Survey (SDSS) \textit{ugriz} photometry 
\citep{Abazajian09,Scott13}, observations with the \SAURON{} IFU
spectrograph \citep{Cappellari11}, 21 cm emission observations \citep{Serra12},
and \textsuperscript{12}CO J=1--0 and J=2--1 observations \citep{Alatalo13}.
These data are available from the \ATLAS{} 
website\footnote{\url{http://www-astro.physics.ox.ac.uk/atlas3d/}}.

The data collected for the \ATLAS{} sample, combined with
extensive dynamical and stellar population modeling, has resulted in a
wealth of valuable measurements. Some relevant observational results include
the presence of optical dust features \citep{Krajnovic11}, surface
brightness profiles \citep{Scott13}, and luminosities  
in the r-band \citep{Cappellari13a}.
We use their stellar population parameters derived from both 
SSP models and reconstructed SFH models 
\citep{McDermid15} to evaluate circumstellar dust models.

Finally, the dynamical analysis and modeling by \citet{Cappellari13a} include 
dynamical mass-to-light ratios derived by fitting model 
\(\langle v_{los}^2 \rangle^{1/2}\) to \(V_{RMS}\) measurements. These were 
derived from Jeans Anisotropic Multi-gaussian expansion (JAM) modeling
\citep{Cappellari08}. The output from these fits include stellar 
mass-to-light ratios derived by 
simultaneously fitting an NFW halo \citep{Navarro96} and a separate stellar 
distribution constrained by the observed surface brightness profile 
\citep{Cappellari13a}.

\section{Data Processing}
\label{sec:processing}

\begin{figure*}
    \centering
    \includegraphics[width=\textwidth]{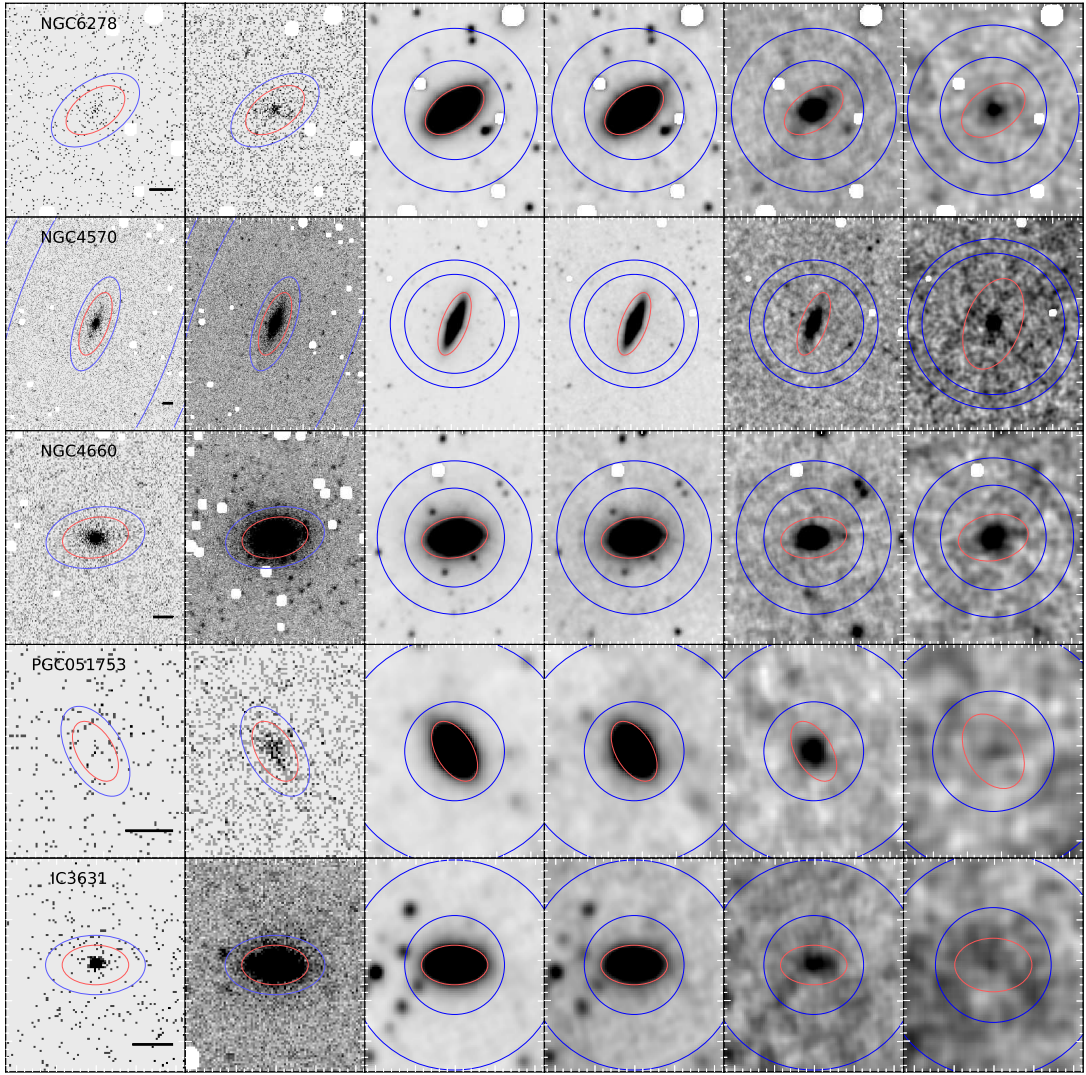} 
    \caption{Cutouts of selected galaxies in \emph{(Left to right:)} \GALEX{} FUV 
    and NUV, and \WISE{} W1, W2, W3, and W4. The white regions are masked 
    pixels, the photometric aperture is in red, and the sky annulus is in blue. 
    Galaxies are shown in order of decreasing \ATLAS{} stellar mass from top to 
    bottom. Scale is preserved from left to right, and a 30\arcsec{} scalebar 
    is shown in the bottom right of each FUV image. IC 3631 is not detected in 
  W4. PGC 051753 is not detected in either FUV or W4.}
\label{fig:cutouts}
\end{figure*}

We chose to perform all aperture photometry
with the standard aperture used by the Two Micron All-Sky Survey (\TM{}) 
Extended Source Catalog
(XSC)\footnote{\url{http://www.ipac.caltech.edu/2mass/releases/allsky/}}: 
the \Ks=20 mag arcsec\(^2\) isophote (hereafter \KT). This isophote 
corresponds roughly to \(1\sigma\) of the typical background noise in the 
\Ks{} images. 
Despite the fact that the \KT{} isophote underrepresents the ``total'' flux 
by \(\sim\mbox{10--20}\%\), it provides the most reproducible measurement of 
a galaxy's
flux\footnote{\url{http://www.ipac.caltech.edu/2mass/releases/allsky/doc/sec4_5a5.html}}.
There are also \TM{} \textit{JH\Ks} measurements of the entire sample in
this same aperture. The sizes of apertures for selected galaxies are shown in
\cref{fig:cutouts}.

Due to PSF differences between the \TM{} and \WISE{} surveys, the effective
shape of the aperture needs to be corrected in order for the isophotes in the
different beams to
match\footnote{\url{http://wise2.ipac.caltech.edu/docs/release/allsky/expsup/sec4_4c.html\#xsc}}.
Because the \WISE{} W4 beam is larger and more circular than the beams
for the other three \WISE{} bands, the W4 aperture is a different
size and shape from the rest. For all but three objects, this adjustment
was performed by the \WISE{} pipeline, which generates a corrected
aperture any time a \TM{} XSC object is centered within 2\arcsec{} of
the \WISE{} source. For the three objects whose centers in the two surveys
are separated by more than 2\arcsec, we generated a matched aperture
manually. We calculated these apertures by binning the \ATLAS{} sample by 
\TM{} axis ratio, and interpolated the adjusted \WISE{} semimajor axis and 
axis ratio to the object. \GALEX{} apertures were not adjusted because the
PSF differences are less important. The 
photometry parameters for all of the galaxies in our sample are given in 
\cref{tab:params}.  

\input{tables/params.tex}

\subsection{\WISE{}}

We measured aperture photometry on the \WISE{} Atlas images. These are
coadded images available as high-level data products. The details of the
image construction and calibration are described in the \WISE{}
explanatory
supplement\footnote{\url{http://wise2.ipac.caltech.edu/docs/release/allsky/expsup/index.html}}.

There are presently two recommended data releases: All-Sky and
AllWISE\@. The AllWISE release is more precise and sensitive because
it incorporates observations from the post-cryo mission into the standard
mission observations, among other improvements. However, inclusion
of post-cryo observations decreases the saturation limit of the images,
and hence decreases accuracy for bright objects. We therefore used
All-Sky images for objects with saturated pixels, and otherwise used
AllWISE\@. The data release of each image is included in \cref{tab:params}.

We measured elliptical aperture photometry with the \ellipse{} package in IRAF 
using the adjusted \KT{} parameters given in \cref{tab:params}. 
Sky values were estimated via the \fitsky{} package in 
IRAF\@.  The inner radius of the annulus was chosen to be 1.5 times 
the semimajor axis of the photometric aperture in order to exclude galaxy flux. 
The thickness of the annulus was chosen to be 30 pixels in order to provide
sufficient sky pixels to adequately measure the background.

We used SExtractor \citep{Bertin96} to detect and mask foreground
sources in the \WISE{} W1 Atlas images, and then applied the masks to
all four \WISE{} bands. 
The most reliable method of foreground removal is PSF subtraction, as
done in \citet{Jarrett13}. However, this
method is not feasible for our large sample of galaxies due to the spatially 
variable PSF in the Atlas images. Since the
stellar emission in ETGs is morphologically smooth, foreground stars
bright enough to significantly affect the galaxy's flux are easily
identified by SExtractor.

\input{tables/mags.tex}

We also analyzed the 17 galaxies (including 3 ellipticals) from \citet{Jarrett13} 
in order to test our  pipeline. We excluded the 
M51 pair because none of our targets require similarly complex deblending.
After comparing the flux measured from our pipeline to those reported by
\citet{Jarrett13}, we encounter RMS differences of W1: 0.05 mag, W2: 0.06
mag, W3: 0.12 mag, and W4: 0.07 mag; we adopt these values as estimates
of our photometric uncertainties. Since the
\citet{Jarrett13} sample is more morphologically complex than our
measurements, we expect these differences are upper limits to the
true photometric uncertainties. These differences cannot be explained
by color corrections, which only correspond to 1\% differences. Our
measurements for the full sample are provided in \cref{tab:magtable}, and have 
not been corrected for extinction. The uncertainties in
\cref{tab:magtable} are formal uncertainties, and include zero-point
uncertainties of 0.006, 0.007, 0.015, and 0.012 mag for W1,
W2, W3, and W4, respectively. For our analysis, we apply extinction
corrections for \TM{} and \WISE{} bands from \citet{Indebetouw05}.

\subsection{\GALEX}

The \GALEX{} GR6/7 data release has images from six different observing 
programs with varying breadths and depths. The deepest is the targeted 
GII program, followed by several science surveys. The surveys vary in
both sky coverage and exposure time, reaching 29,000~s for fields with
well-studied galaxies, down to \(\sim100\)~s for the shallowest survey,
which covers 63\% of the sky. For galaxies observed in multiple
surveys, we chose the highest exposure-time image which contained the
entire photometric aperture in the field-of-view. Although the PSF is
distorted near the edge of the field \citep{Morrissey07}, we
found that this did not affect the photometry. The tilename for each
galaxy identifies the exposure and is given in \cref{tab:params}.

We similarly used \ellipse{} in IRAF to measure the \GALEX{} images,
although we estimated the sky differently. While the \GALEX{} pipeline 
provides background
images, we noticed galaxy flux in the background images for some
objects. We therefore opted for traditional background estimation from
an annulus.  Unfortunately, \textit{fitsky} in IRAF does not perform
well when the background counts are very small, as it expects sky values
to be normally distributed. We therefore
implemented the method described in \citet{GildePaz07}, which divides two 
elliptical annuli into radial segments and averages over those segments.  
Similar to \citet{GildePaz07}, we divided the annuli into
90 segments total, with a typical segment area of 4000 
pixels. We set the semimajor axis of the inner annulus at 1.5 times the
semimajor axis of the photometric aperture. The 
uncertainty of the background comes from the standard deviation of the
segment mean values.

We tested our approach with 20 morphologically
diverse galaxies from the \citet{GildePaz07} atlas which overlapped 
our sample, and used the same D25 aperture. Despite
using the same technique, we systematically measured less
flux than \citet{GildePaz07} by 0.1 to 0.4 mag. We did
successfully reproduce the
\citet{GildePaz07} results with the original cutouts from the NASA/IPAC
Extragalactic Database (NED)\footnote{\url{https://ned.ipac.caltech.edu}}, so 
we conclude that the discrepancy is due to changes in the \GALEX{} pipeline.
We also compare our measurements with \citet{Bai15}, who
remeasured the \citet{GildePaz07} atlas using data processed by the current 
GR6/7 pipeline. Our results agree with \citet{Bai15} to an RMS difference of 
0.14 mag without a systematic trend in both the NUV and FUV bands, which 
we attribute to different background estimation methods. We 
conclude that our UV measurements are accurate and have an RMS precision of 
0.14 mag.

Our \GALEX{} measurements are also in \cref{tab:magtable}. They have not
been corrected for extinction, although for all of our analysis
we used the prescription in \citet{GildePaz07} to apply extinction
corrections to these measurements. Values for E(B-V) are from the
\citet{Schlegel98} maps obtained from the IRSA dust map service.

\subsection{Separating Dusty Galaxies}
\label{sec:contamination}

\begin{figure}
    \centering
    \includegraphics[width=\columnwidth]{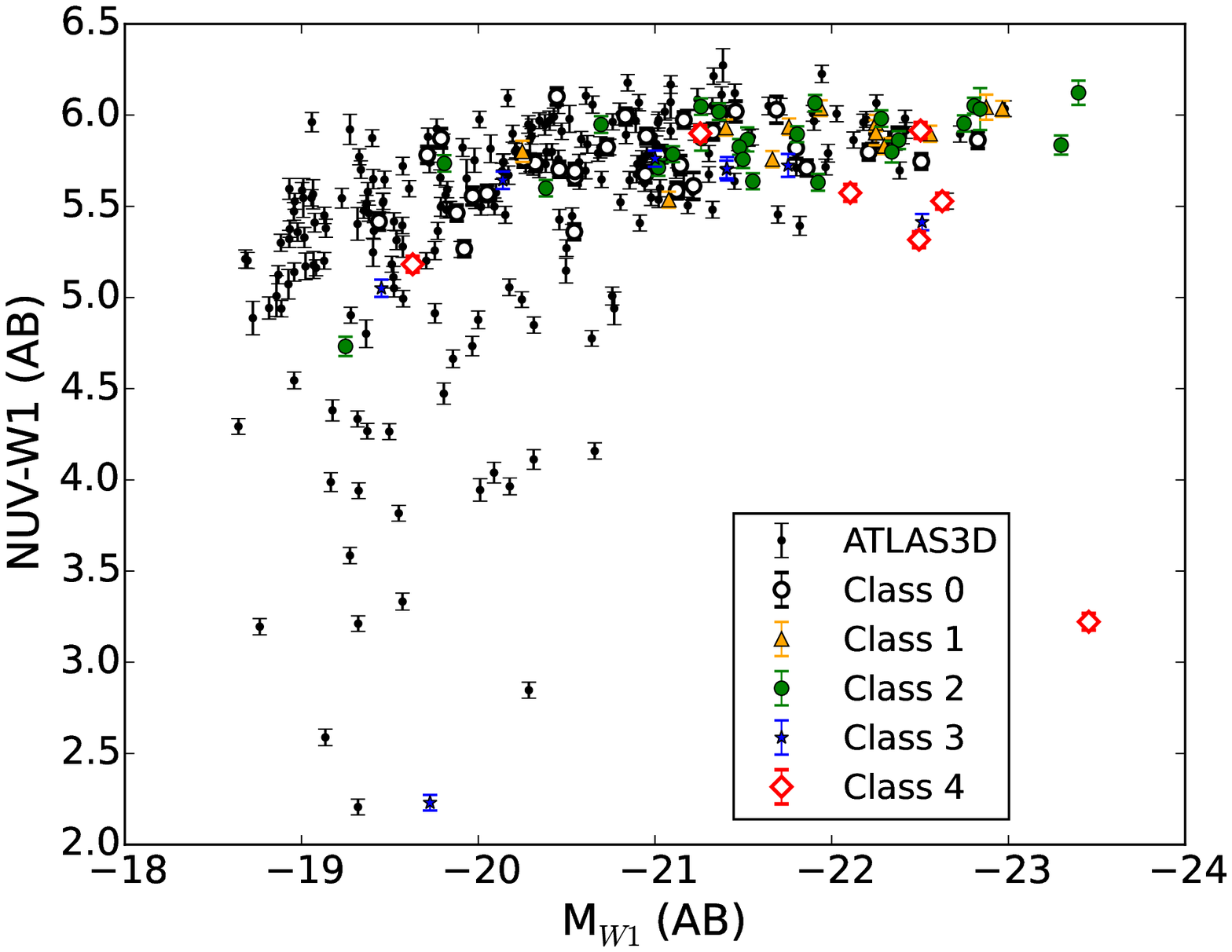}
    \caption{NUV-W1 vs W1 luminosity for the \Rampazzo{} and \ATLAS{}
    galaxies. These quantities are reasonable proxies for the
shape of the radiation field and stellar mass, respectively. There is no
clear trend between \Rampazzo{} class and NUV-W1, even for the \Class{2}
galaxies with anomalous PAH ratios. The luminous, blue \Class{4}
  object is NGC 1275, which is a Seyfert 1.5 and thus may not be
representative of ETGs.}\label{fig:nuvw1}
\end{figure}

\Cref{fig:nuvw1} shows a color-absolute magnitude diagram that
illustrates the diversity in the shape of the SED
for the \ATLAS{} and \Rampazzo{} galaxies.
The majority of ETGs are extremely deficient in UV photons, as is
expected from their generally old stellar populations. There appears to be a 
trend in the SED,
where more UV-rich galaxies tend to be less luminous. This trend appears
unrelated to the MIR classes of \Rampazzo{}, although the RSA \IRS{} atlas
contains few low-luminosity ETGs. The region with blue ETGs is 
also sparely populated compared to the area with NUV-W1\(>5\). The
\Class{4} object which is an outlier at the bottom-right is
NGC 1275, a Seyfert 1.5 galaxy in the \Rampazzo{} sample.
There does not appear to be foreground contamination for this galaxy, so
we believe the extreme NUV-W1 color and luminosity are due to the AGN
component.

As indicated by \citet{Rampazzo13}, about half of ETGs contain observable
traces of interstellar dust, which can dominate the MIR signal. We
therefore attempted to separate the passive from the non-passive galaxies.
For the \Rampazzo{} sample, non-passive galaxies are
classified from \IRS{} spectra as \Class{1--4} fairly reliably. The
\ATLAS{} sample does not have MIR spectra, so we attempt to distinguish
between passive and non-passive galaxies with the extensive ancillary data.

\citet{Martini13} demonstrated a one-to-one correspondence between
optical dust lanes observed with the \textit{Hubble Space Telescope
(HST)} and emission from cold dust detected by \Spitzer{} MIPS\@. We attempted 
to remove obvious contaminants with observed dust features in r-band
observations \citep{Krajnovic11}. However, since the photometric
resolution of SDSS is significantly lower than HST, we also searched
for false negatives in our sample by cross-matching the dust detections
in \citet{Martini13} to the non-detections in \citet{Krajnovic11}. This
revealed many cases where the \ATLAS{} images did not reveal dust lanes
that were clearly visible with HST and via FIR detections, so we also excluded
galaxies with CO detections \citep{Young11}. 

\begin{figure}
    \centering
    \includegraphics[width=\columnwidth]{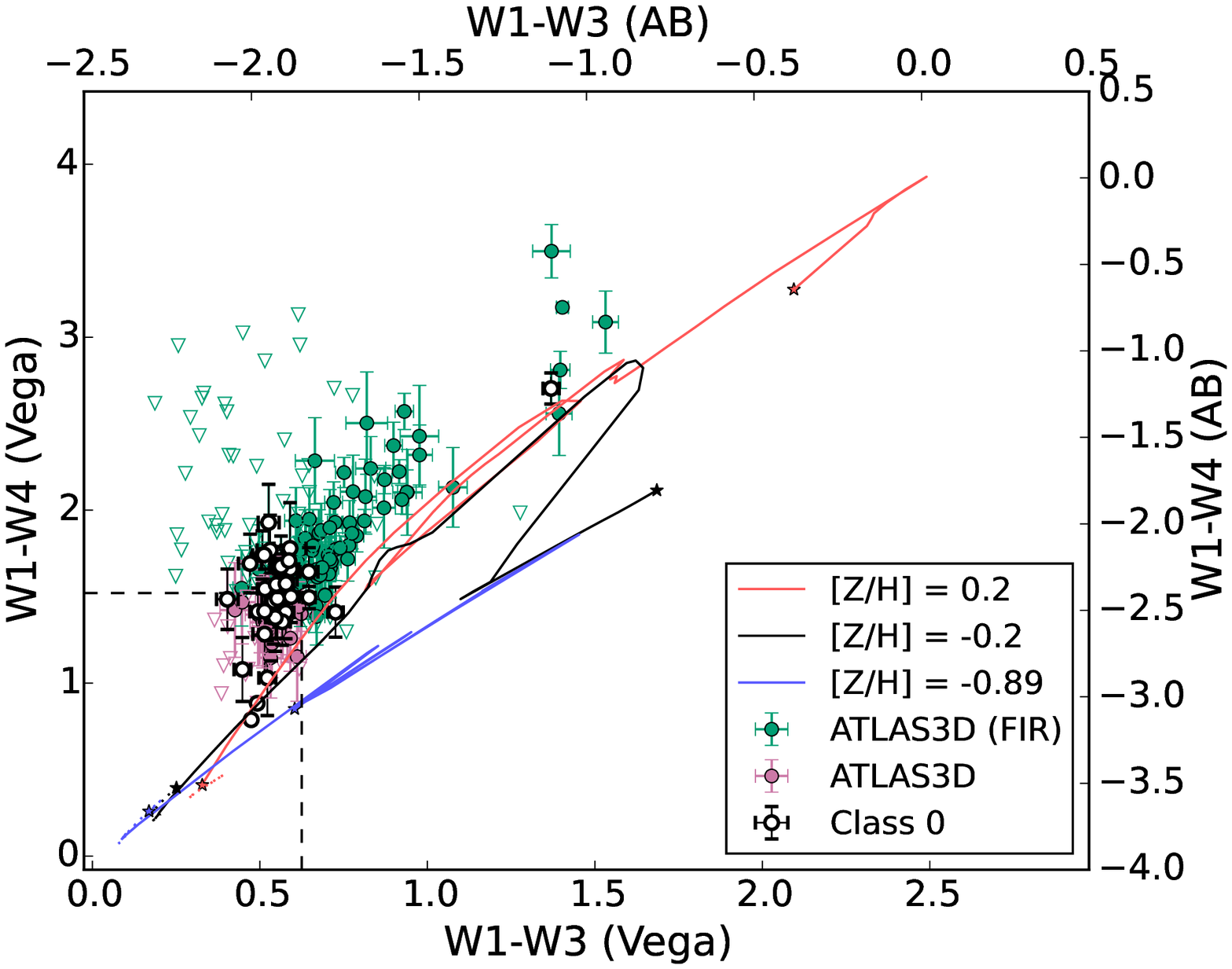}
    \caption{W1-W3 vs. W1-W4 colors of \ATLAS{} galaxies without CO or
    dust detections. Circles are data and triangles are upper limits
    due to W4 non-detections. The dashed black lines represent the
    ``color-cut'' criteria. Magenta and green points represent
    galaxies inside and outside the ``color-cut'' dustless region,
    respectively. Also shown is the passive \Class{0} sample from
    \citet{Rampazzo13}. These data are compared to circumstellar
    tracks from Flexible Stellar Population Synthesis models. The red 
    track corresponds to the highest metallicity galaxy in the \ATLAS{} sample 
    with [Z/H]=0.2. The black track corresponds to the median metallicity in 
    the \ATLAS{} sample of [Z/H]=-0.2.
    The blue track corresponds to the lowest metallicity galaxy with
    [Z/H]=-0.89. The beginning of the track at 1 Gyr 
    is marked with a star, and continues until 14 Gyr. Solid tracks 
    represent the FSPS model with the circumstellar dust model of
    \citet{Villaume15}; the very short dotted tracks in the lower left corner 
    are the same models without circumstellar dust.}\label{fig:dustless}
\end{figure}

\Cref{fig:dustless} shows the W1-W3 vs. W1-W4 colors for galaxies without 
evidence for diffuse, interstellar dust.  We also compare this sample to SPS 
tracks with metallicities that bracket the \ATLAS{} sample. We used SPS models
with an exponential SFH timescale of 100 Myr. We also included a minimum sSFR of
\(10^{-14}\) yr\(^{-1}\) to account for constant, very low
levels of star formation \citep{Ford13}. \cref{fig:dustless} shows that the 
Flexible Stellar Population Synthesis (FSPS) high and median metallicity
model tracks follow the shape of the data quite 
well, albeit with an offset. We explore the source of this offset in 
\cref{sec:circdust}. 

When compared to the passive \Class{0} galaxies from \citet{Rampazzo13},
there still appears to be a tail extending redward of the clump of
\Class{0} galaxies. The single \Class{0} object in the tail is NGC 4377, 
which has potential foreground contamination. Therefore, we performed an
external check on this sample by cross-matching with \Herschel{} detections
at FIR wavelengths \citep{Smith12,diSeregoAlighieri13,Amblard14}. For objects in
\citet{Amblard14} we defined a ``dust detection'' as a 5\(\sigma\) detection in 
at least one of the 250 \micron, 350 \micron, and 500 \micron{} bands. 
For galaxies from the other two studies, we used their internal criteria 
to indicate a dust detection. We note that two of the 
\Class{0} objects had \Herschel{} FIR detections, along with 31\% of the 
overlapping \ATLAS{} sample, which suggests there is still substantial 
contamination. The region of the color-color diagram not populated by the 
galaxies with FIR detections is \(W1-W3 < 0.67\) and \(W1-W4 < 1.52\) (AB\@: 
\(W1-W3 < -1.85\) and \(W1-W4 < -2.40\)). NGC 4486A, a tidally disrupted
satellite of M87, is the only galaxy
in this region with a marginal \Herschel{} detection at 250 \micron. Because 
of its tidal interactions, we classify it as
a peculiar case, and assume that the rest of the galaxies in this region do not contain interestellar dust. We denote the galaxies in this 
region as the ``color-cut dustless'' sample, compared to the subset with 
only CO and/or dust exclusions, which we just term ``dustless''.

\section{Stellar Population Synthesis Models and Circumstellar Dust}
\label{sec:emission}

\begin{figure}
    \centering
    \includegraphics[width=\columnwidth]{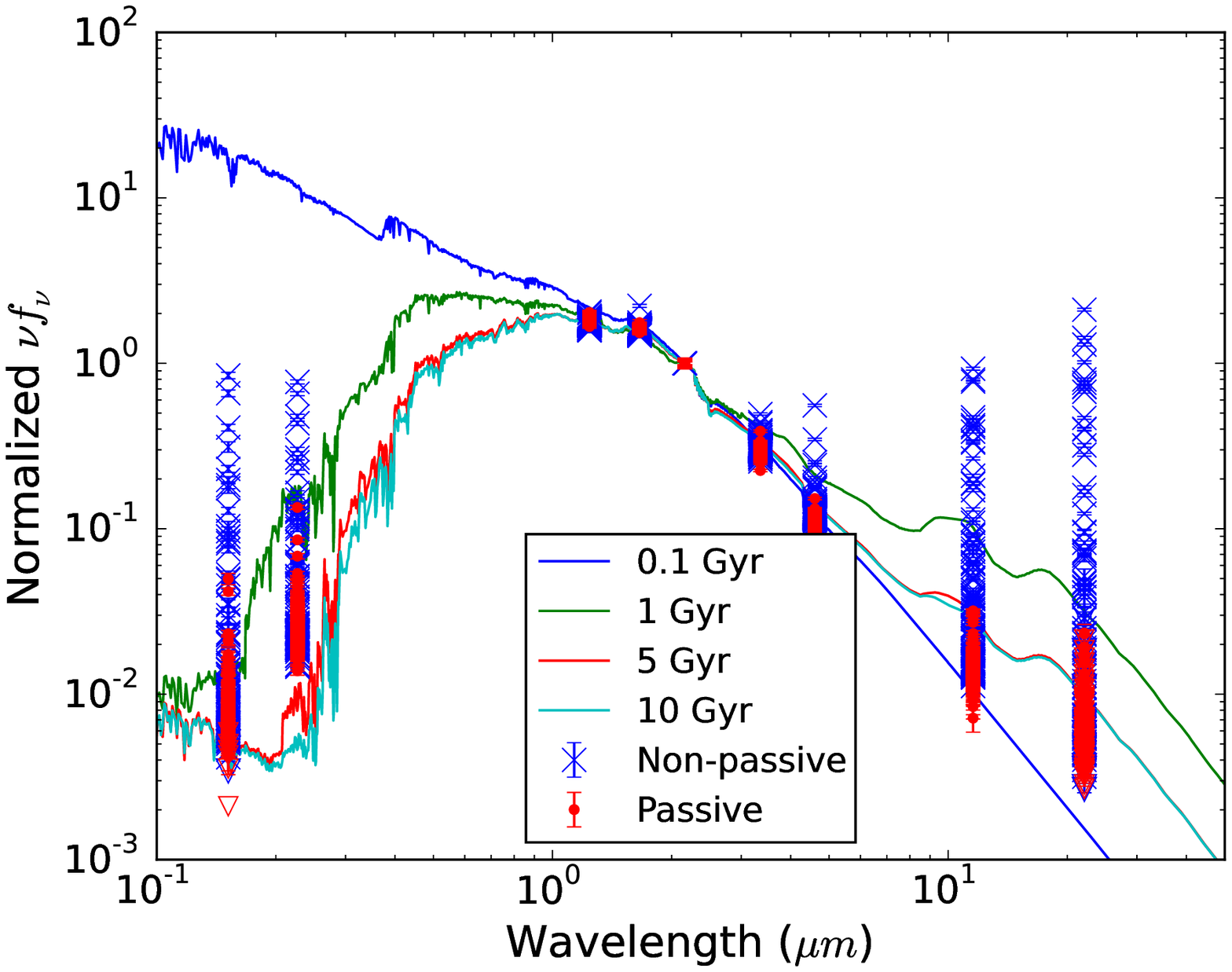}
    \caption{Spectral Energy Distributions of galaxies within the \KT{}
    aperture from broadband FUV, NUV, J, H, Ks, W1, W2, W3, W4
    photometry, normalized to the \Ks{} band. Model spectra are
    high-metallicity FSPS models \citep{Conroy09,Conroy10} with exponentially 
    decreasing SFH with 
    \(\tau_{SFH} = 100\) Myr and [Z/H] = 0.2. The different spectra
    represent t=0.1, 1, 5, and 10 Gyr ages. Data points are the combined 
    \ATLAS{} and \Rampazzo{} galaxies. \ATLAS{} galaxies without CO
    detections and dust lane morphology, or \Rampazzo{} \Class{0}
    galaxies make up the passive group,
    while those with either detection, or classified as
    \Rampazzo{} \Class{1--4}, are in the non-passive group.}\label{fig:seds}
\end{figure}

We used the \GALEX{}, \TM{}, and \WISE{} photometry to construct broadband
SEDs for our sample. These SEDs are shown in \cref{fig:seds}, normalized to the
\Ks{} band. NGC 2974 is a very large outlier in W1 and
W2, most likely due to very severe foreground contamination from a nearby
bright star. While its photometry is reported in \cref{tab:magtable},
we exclude it from our plots due to its suspect color. 

We compare our data to the FSPS v2.5 models \citep{Conroy09, Conroy10}, which 
recently incorporated circumstellar dust \citep{Villaume15}.
These models allow for very extensive customization of isochrones and
stellar atmosphere libraries. The FSPS models match the data well, 
particularly in the UV\@. In W3 and W4 many of the points extend above the 
model predictions. However, galaxies in the dustless sample described in
\cref{sec:contamination} fall well within the range of
colors predicted by the models.

\subsection{Stellar Mass-to-Light Ratios}
\label{sec:masstolight}

One valuable use for SPS models has been to predict stellar
mass-to-light ratios (M/L) for galaxies from broadband colors 
\citep{Bell01,Bell03,Zibetti09}.  \citet{Bell01} illustrated the power of 
this approach with their relation between \(M/L_{B}\) and \(B-R\) color, which
was largely robust to metallicity, extinction, and bursty star formation
history (SFH).
More recent improvements to M/L determinations include the use of multiple 
colors and models of dust extinction \citep{Zibetti09}. Yet despite these
improvements, the presence of young stellar populations and internal 
extinction remains a major source of uncertainty for these models
\citep{Bell01,MacArthur04,Zibetti09}.  
These uncertainties are less severe at longer
wavelengths, where the effects of young stellar populations and
extinction on the mass-to-light ratio are diminished \citep{Meidt14}.
However, this reduction in uncertainty comes at a cost of increased 
uncertainties in modeling the Asymptotic Giant Branch (AGB) phase in the
stellar models, which are especially poorly studied at low metallicity
\citep{Maraston06}. SPS models predict that the impact of AGB stars peak around 
1 Gyr and decrease at greater ages \citep{Melbourne12}, so should be less
important for most ETGs.

Many different prescriptions exist to determine stellar masses of galaxies
\citep[e.g.][]{Meidt14, Cluver14} that are either directly based on or calibrated
from SPS models \citep{Bruzual03,Salim07}.
However, the accuracy of these prescriptions is limited by the
uncertainties associated with models for evolved stars, the dust geometry, 
IMF, and SFH.

Stellar mass-to-light ratios determined by dynamical means from
\ATLAS{} are extremely
valuable because they are not dependent on the same assumptions made by SPS 
models, such as an IMF\@. Results from \ATLAS{} indicate a tension between
stellar masses generated from SPS models and from dynamics, which is
correlated with observed velocity dispersions. This
disagreement has been put forward as evidence that the IMF may be 
variable \citep{vanDokkum12, Cappellari12}, which would imply greater and 
systematic uncertainties in SPS-derived stellar masses that assume a 
universal IMF\@. 

\begin{figure}
    \centering
    \includegraphics[width=\columnwidth]{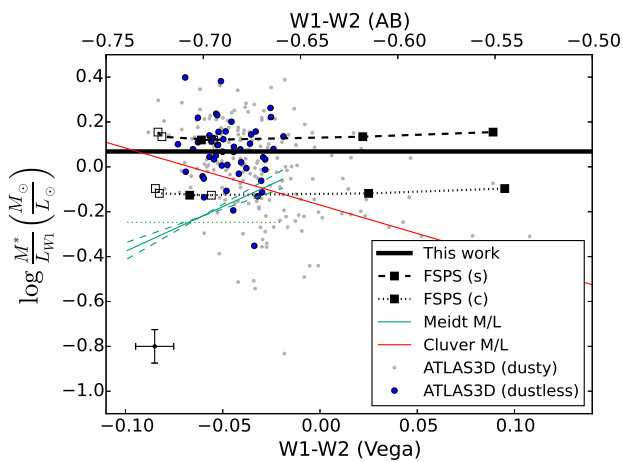}
    \caption{Stellar mass-to-light in the W1 band for the \ATLAS{} 
        sample along with several SPS predictions and empirical
        measurements. The data points are converted from the
        \ATLAS{} r-band stellar mass-to-light ratio, corrected to 
            match the \TM{} K20 aperture. Large blue points meet the
            color-cut dustless criteria defined in
            \cref{sec:contamination}, while small grey points do not. A
            representative error bar is given in the lower left.
            The red line is the relation of \citet{Cluver14} from
            the GAMA survey. The green solid and dashed lines indicate the 
        M/L vs. W1-W2 relation and dispersion from \citet{Meidt14} for
        old, stellar populations, derived from the \citet{Bruzual03} models. 
        The horizontal dotted green line represents the color-independent M/L 
        suggested by \citet{Meidt14} using an age-metallicity relation for ETGs.
      The dashed and dotted lines connect square points from FSPS models with
      \Salpeter{} and \Chabrier{} IMFs, respectively,  evaluated at
    three different metallicities (from left to right): [Z/H] = -0.89, 0.0, 
    0.20 for a 10 Gyr population. Closed squares use W1-W2 directly from FSPS\@. 
    Open squares use W1-W2 colors as corrected by \citet{Meidt14}.}\label{fig:masstolight}
\end{figure}

In order to determine if the tension is also seen in the MIR, we transformed 
the \ATLAS{} mass-to-light ratio from r-band to W1, and corrected for the 
difference between the largest Multi-Gaussian Expansion (MGE) aperture used in 
\ATLAS{} and the modified \KT{} aperture used here. The band was transformed 
according to the formula:
 \begin{equation}
   \begin{split}
   \log_{10} & \left(\frac{M}{L_{W1}}\right)_{\KT{}} = \log_{10}
     \left(\frac{M}{L_r}\right)_{MGE} + \log_{10}
     \left(\frac{L_r}{L_{r,\sun}}\right) \\
     &+ 0.4 \left(\left.m_{W1}\right|_{MGE} - 5
         \log_{10} \left(\frac{d}{10 pc}\right) - M_{W1, \sun}\right)
   \end{split}
 \end{equation}
where \((M/L_{W1})_{\KT}\) is the mass-to-light ratio within the
\KT{} aperture in W1, \((M/L_r)_{MGE}\) is the \ATLAS{} mass-to-light
ratio within the region where the surface brightness is accurately
modeled by the MGE, \(L_r/L_{r,\sun}\) is the \textit{r}-band luminosity
of the galaxy in units of the in-band solar luminosity with \(M_{r,
\sun} = 4.64\) in AB magnitudes \citep{Blanton07}, \(m_{W1}\) is the AB 
magnitude of the galaxy as measured in this work, \(d\) is the distance to 
the galaxy, and \(M_{W1, \odot}=5.94\) mag is 
the absolute magnitude of the Sun in AB magnitudes (transformed from the Vega 
value in \citet{Jarrett13}). This 
transformation assumes that the stellar mass-to-light ratio is spatially 
constant, which is consistent with assumptions made in JAM
modeling \citep{Cappellari13a}. In order to find an
aperture representative of the total flux of the galaxy 
measured by \ATLAS{}, we numerically integrated the MGE intensity model in an
elliptical aperture such that at least 90\% of the flux would
be in that aperture. This cutoff was chosen because the accuracy of the
integrated flux of the MGE models compared to SDSS fluxes is 10\%
\citep{Scott13}. The transformed mass-to-light ratios are shown in
\cref{fig:masstolight}. The uncertainty of the mass-to-light ratio for
each data point includes the photometric and zero-point uncertainties in W1, 
the photometric uncertainty in r-band \citep{Scott13}, JAM modeling 
uncertainties \citep{Cappellari13a}, and distance uncertainties 
\citep{Cappellari11}.  

Although extinction is greatly reduced in the MIR, nonstellar emission---dust 
in particular---can be a major contaminant, introducing uncertainties of up to 
30\% when integrated over the entire galaxy \citep{Meidt12}. We
therefore use the color-cut dustless sample, described in
\cref{sec:contamination}, to minimize interstellar dust contamination.

The color-cut dustless galaxies in \cref{fig:masstolight} do not appear
affected by diffuse, interstellar dust, nor is there a
noticeable trend in W1-W2. We calculate that \(\log_{10} M/L_{W1} = 0.07 \pm 0.13\) for old 
stellar populations without diffuse dust. This value contrasts
with the result of \citet{Meidt14} of \(\log_{10} M/L_{W1} = 0.25 \pm 0.11\).
The offset is consistent with the IMF transition from \Chabrier{} to 
\Salpeter{} with increasing velocity dispersion found by 
\citet{Cappellari12} and \citet{vanDokkum12}. To
illustrate the continuum of IMFs, we include FSPS tracks along metallicities
\([Z/H] = -0.89\), \([Z/H] = 0.0\), and \([Z/H] = 0.20\) which include the 
bounds of the \ATLAS{} sample and solar metallicity, with both \Chabrier{} and 
\Salpeter{} IMFs. The majority of objects, as well as the average 
\(M/L_{W1}\), lie between these two bounds. This explanation fits 
our observations better than the \Chabrier{} IMF assumed by
\citet{Meidt14}. 

The data may have a smaller range of W1-W2 than predicted by the FSPS
models because the FSPS models do not include the metallicity-dependent
CO absorption line in W2 \citep{Peletier12,Norris14}. We applied the
GLIMPSE-calibrated correction described in \citet{Meidt14} to the W1-W2
colors, which resulted in a narrower and slightly bluer range of W1-W2 than 
observed in the color-cut dustless sample.

One factor which may affect our interpretation of the M/L difference
is that the color-cut used to separate dustless galaxies could
preferentially select against galaxies with certain velocity
dispersions if there is a trend with stellar populations. We found that the 
color-cut did preferentially select high velocity dispersion galaxies, so we 
fit a second time and only excluded galaxies with CO/dust detections. The new 
criteria sampled a broader range of velocity dispersion and resulted in a 
revised value of \(\log M/L_{W1} = -0.03 \pm 0.17\). This is closer to the
\citet{Meidt14} value, but still discrepant at the \(2\sigma\) level.

We also compare our results to an empirically-derived M/L ratio from GAMA
using \WISE{} colors \citep{Cluver14}. This relation differs from that
in \citet{Meidt14} in two respects. First, the
\citet{Meidt14} relation was fit to a grid of \citet{Bruzual03} models,
rectilinear in age and metallicity. In contrast, the \citet{Cluver14}
relation was fit to the resolved portion of the GAMA sample, with
stellar masses derived from optical colors \citep{Taylor11}. This
difference in how the fits are populated likely leads to the difference
in the slope. Secondly, the \citet{Meidt14} relation was fit to a pure 
stellar population, while the \citet{Cluver14} relation was fit to
very dusty galaxies. This likely explains why the relation seems to 
trace the dusty population, but is very discrepant with the dust-free ETGs. 

\subsection{Evidence for Ubiquitous Circumstellar Dust}
\label{sec:circdust} 

All low- to intermediate-mass stars (between 0.5--8 M\(_{\sun}\)) pass through
the Asymptotic Giant Branch (AGB) when undergoing hydrogen and helium shell
burning \citep{Marigo08}. These stars have extremely cool and tenuous atmospheres where dust 
grains condense and drive mass-loss 
\citep{Salpeter74a, Salpeter74b, Goldreich76, Bedijn87}. As a result, we expect
to observe this stage of stellar evolution in populations with ages ranging
from 100 Myr to greater than the age of the Universe. However, SSP
models indicate that the effects of circumstellar dust on integrated MIR flux 
peaks at 1 Gyr after star formation, and then decreases significantly by
10 Gyr \citep{Villaume15}.

Circumstellar dust emission has been proposed as an age tracer of old
stellar populations because broadband optical colors have an age-metallicity
degeneracy \citep{Bressan98}. Despite the lack of
specific MIR age indicators for the \WISE{} bands, SPS modeling can 
potentially yield an evolutionary track for a given color due to 
circumstellar dust \citep{Villaume15}. \citet{Bregman06} showed that
there is some tension between ages determined from optical line indices
and circumstellar dust fits. We compare these two
age determinations for the \ATLAS{} sample using the FSPS
models. We also investigate the effect of circumstellar dust on star formation
rate indicators in the MIR for galaxies with low specific SFRs (sSFR = SFR / 
\(M_*\)).

We used MIR colors to obtain a distance-independent measurement of 
circumstellar dust. Flux in the short-wavelength W1 and W2 bands is
largely dominated by the stellar continuum and is not sensitive to dust
\citep[e.g.][]{Villaume15}. In contrast, the longer-wavelength W3 and W4 bands should be
increasingly sensitive to dust, since they lie in the region of the
spectrum where circumstellar dust emission dominates over photospheric
emission. We avoid the W2 band in comparing data to models because the
effect of CO absorption alluded to in \cref{sec:masstolight} has not been
included in all SPS models \citep{Peletier12,Norris14}. 

\begin{figure}
    \centering
    \includegraphics[width=\columnwidth]{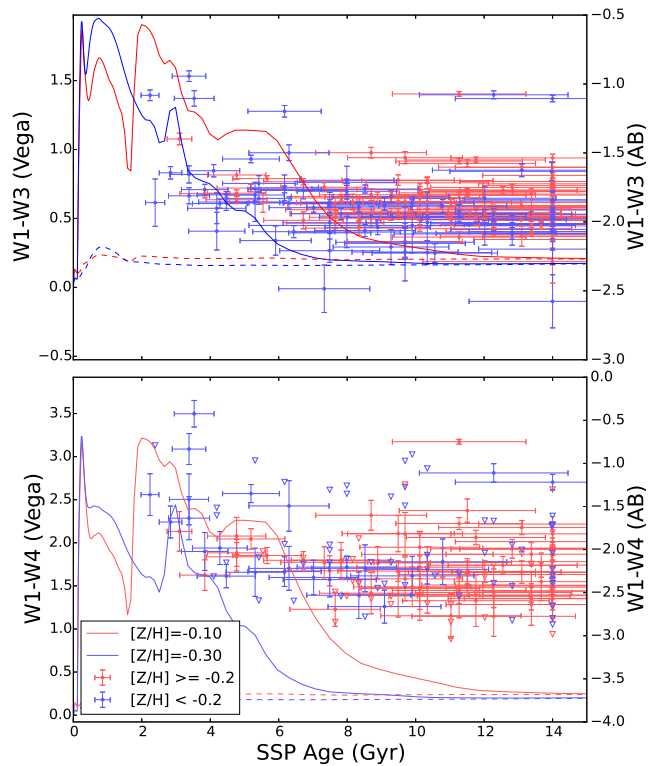}
    \caption{Colors of \ATLAS{} galaxies vs.\ SSP age compared with color 
    evolution from FSPS for W1-W3 \emph{(Top)} and
    W1-W4 \emph{(Bottom)}. Red and blue lines represent SSP models 0.1
    dex above and below the median metallicity of the \ATLAS{} sample,
    which should bound 60\% of the \ATLAS{} sample in metallicity.
    The solid lines represent the color evolution of the FSPS model with the 
    circumstellar dust prescription of
\citet{Villaume15}, while the dashed lines represent the same population
without circumstellar dust. The \ATLAS{} points are similarly
color-coded and represent galaxies which lie above and below the median
metallicity value of \([Z/H] = -0.2\). Galaxies with SSP ages greater
than 14 Gyr are reset to 14 Gyr. For objects without a robust W4
detection, upper limits to the color are shown instead.}
    \label{fig:cdust}
\end{figure}

\begin{figure}
    \centering
    \includegraphics[width=\columnwidth]{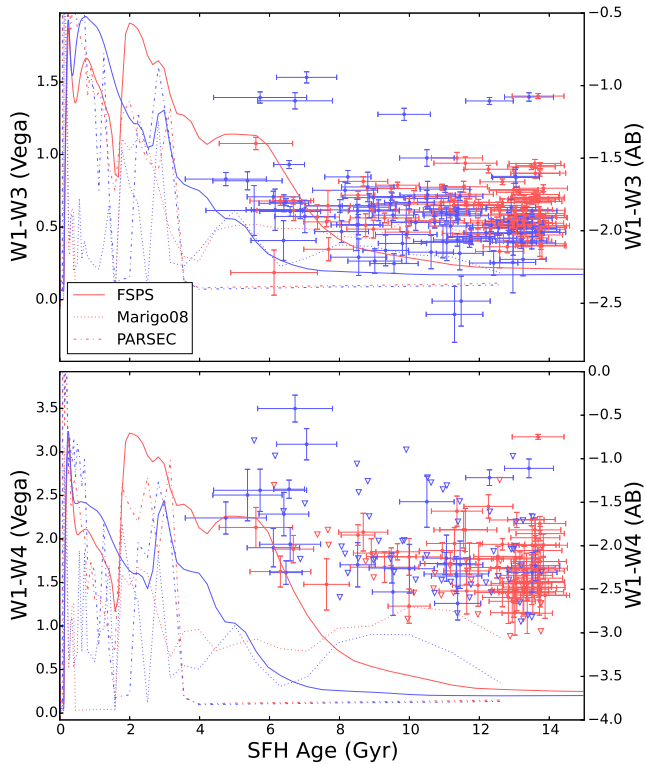}
    \caption{Comparison between models and data similar to \cref{fig:cdust}, 
    except with SFH ages instead of SSP ages. Red and blue colors have
    the same meaning as in \cref{fig:cdust}. Line styles correspond to 
    different SPS models: solid lines correspond to FSPS models, dotted lines 
    correspond to \Marigo{} SSP models, and dot-dashed lines correspond to the 
    PARSEC SSP models.}
    \label{fig:sfhcdust}
\end{figure}

It is apparent from \cref{fig:dustless} that circumstellar dust is
necessary in order to explain MIR colors; however, there is an offset
between the tracks and the data. We
explore the source of this offset in more detail in
\cref{fig:cdust,fig:sfhcdust}. In \cref{fig:cdust} we used SSP ages for the 
\ATLAS{} sample from \citet{McDermid15} derived by
fitting optical line indices. Since the SSP ages in \citet{McDermid15}
were not constrained to be consistent with the age of the Universe, we
set the galaxies with SSP ages greater than 14 Gyr to have ages of 14
Gyr, in order to be consistent with current cosmological models
\citepalias{Planck14}; this was 19\% of the sample. While the discrepancies 
in W1-W3 do not seem be too 
large, the models appear to underpredict the flux in W4 by a factor of 2.5, 
which is comparable to the offset seen in \cref{fig:dustless}. As seen in 
\cref{fig:sfhcdust}, mass-weighted ages estimate galaxies to be older,
making the offset even more apparent. The potential causes of this
discrepancy could include a ubiquitous intermediate-age population whose 
influence is seen in the MIR, but not detectable in the optical, and that a 
wider mass range of AGB stars produce circumstellar dust at subsolar 
metallicities than predicted by the models. 

\cref{fig:sfhcdust} compares the FSPS models to the PARSEC v1.2S + COLIBRI PR16 
SSP models \citep{Bressan12,Marigo13,Rosenfield16} and the \Marigo{}
models. The PARSEC models seem to predict no circumstellar dust 
at late ages, exacerbating the tension seen with the FSPS models. On the other 
hand, the \Marigo{} models predict more dust at late times compared to both the 
FSPS and PARSEC models. However, the predicted colors of the dusty stellar 
population are still bluer than indicated by the data.  

While some of the extremely red galaxies are likely contaminated by 
interstellar dust, it is significant that very few galaxies are consistent with 
having no circumstellar dust. Since the number of galaxies redder than the
no-circumstellar-dust models is significantly greater than our estimated 
contamination rate, we conclude that circumstellar dust is the source of the 
MIR excess for galaxies without interstellar dust.  These results are robust to
differences in the IMF. 

\subsection{Impact of Circumstellar Dust on MIR SFRs}

Accurate Star Formation Rates (SFRs)
provide valuable insights into galaxy formation. The most direct method
to determine SFR measures Balmer emission from HII regions, which is
related to the number of ionizing photons from young stars; however, the
extinction corrections can be substantial and uncertain \citep{Kennicutt98a}. 
SFRs derived from MIR to FIR observations  are more indirect as they measure 
the SFR through radiation reprocessed by dust grains. This method is quite 
insensitive to extinction uncertainties, but
at low SFRs the old stellar population may also significantly heat the
dust \citep{Helou86, Lonsdale87, Kennicutt98b, Groves12}. 

We quantify the important contribution of circumstellar dust to the total
MIR luminosity, and its impact on SFR estimates with \WISE{} W1, W3 and W4. 
W1 traces the old stellar population that dominates 
stellar mass, and W3 and W4 trace the PAHs and warm dust 
heated by young stars, respectively, so sSFR can be calculated
without additional observations. Although W4 is considered a superior
tracer of SFR over W3, fainter galaxies are more readily detected in W3.
Since SFRs can be measured in W3 for galaxies too faint for a W4
detection, we include it in our analysis of SFR measurements.

\citet{Davis14} investigated the impact of circumstellar dust on SFR for
the \ATLAS{} sample using \WISE{} catalog photometry, with prescriptions
calibrated for \Spitzer{} bands. We reinvestigate this prescription with our 
updated photometry, more stringent criteria for dustless galaxies, and with 
the more recent SFR relations of \citet{Cluver14}.

\begin{figure}
  \centering
  \includegraphics[width=\columnwidth]{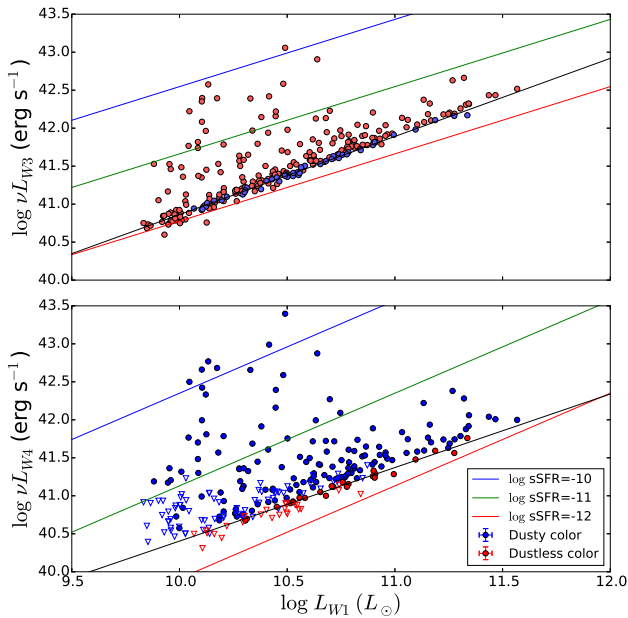}
  \caption{\emph{(Top)} W3 vs. W1 and \emph{(Bottom)} W4 vs. W1 luminosities 
  for the \ATLAS{} sample. Circles
    are detections and triangles are W4 upper limits. Red
  points are in the dustless color-cut sample. Error bars are
  smaller than the points. The blue,
green, and red lines represent where specific star formation rate
is \(10^{-10}\) yr\(^{-1}\), \(10^{-11}\) yr\(^{-1}\), \(10^{-12}\)
yr\(^{-1}\). The black line represents a fit to the color-cut dustless galaxy
detections. Note that W3 is significantly more sensitive than W4.}
    \label{fig:circsfr}
\end{figure}

We plot the W1 vs. W3 and W4 luminosity for the \ATLAS{} sample in 
\cref{fig:circsfr} 
(see Figure 1 of \citet{Davis14} for a similar plot with \Ks{}). 
The color-cut dustless sample described in \cref{sec:contamination} is shown in 
red. The color-cut sample was chosen because it should have less
interstellar dust contamination compared to the regular dustless sample, 
and should represent an accurate relation between W1 and W3 (or W4) in the 
absence of star
formation. Because we may have excluded some truly dust-free galaxies
whose colors do not match our exclusion criteria, the scatter in our
relation may be underestimated. Follow-up observations of the \ATLAS{} sample 
with FIR/submillimeter measurements or high-resolution
visible-wavelength images to search for dust lanes would result in a more 
representative sample of dustless galaxies. 

In order to characterize the minimum SFR which can be usefully measured with W4
observations, we calculate sSFR limits with the relations in
\citet{Cluver14} and use the mean W1-W2 value of our sample to
get \(M/L_{W1}\). The results shown in \cref{fig:circsfr} indicate that
the contribution of circumstellar dust to W3 and W4 will mimic an sSFR of \(2 
\times 10^{-12}\) yr\(^{-1}\).  Lower values of sSFR cannot be reliably 
inferred with integrated \WISE{} photometry alone.

Even at larger sSFR values, it is still necessary to remove the effect
of circumstellar dust. In order to quantify the effect, we fit a line to 
the dustless detections and obtain:
\begin{align}
  \log \nu L_{W3} &= 1.03 \log L_{W1} + 30.58, \\
  \log \nu L_{W4} &= 0.97 \log L_{W1} + 30.72
\end{align}
with an RMS scatter of 0.03 and 0.04 dex, where \(L_{W1}\) is the
``in-band'' luminosity in solar luminosities, and \(\nu L_{\nu}\) is the
spectral luminosity of the band in erg s\(^{-1}\). These relations may be used 
to subtract the circumstellar dust contribution to W3 or W4 for relatively
quiescent galaxies. The W4 relation is similar to the relation derived by 
\citet{Davis14} between Ks and W4.  These corrections will likely be 
negligible for typical star-forming galaxies as the sSFR of local \(L^*\) 
galaxies is around \(10^{-10}\) yr\(^{-1}\) \citep{Cluver14}. 

\section{PAH Ratios with \GALEX{} and \WISE{}} 
\label{sec:pahs}
 
Mid-infrared spectra of early-type galaxies have shown that many exhibit
much weaker short-wavelength PAH features (6.2, 7.7, and
\eightsix{}) relative to those at longer wavelengths
(e.g.\ 11.3 and \twelveseven{})
\citep{Kaneda05,Kaneda08,Vega10,Rampazzo13} compared to star forming galaxies 
\citep{Helou00,Brandl04,Smith07}. Observations of later-type galaxies have also shown 
that low values of these band ratios are commonly found in low-luminosity AGN 
\citep{Smith07,Odowd09}. Many early-type galaxies contain evidence for LINERs 
\citep{Ho97}, although as most may not be dominated by AGN
\citep{Sarzi10}, it is not clear if there is a direct connection between AGN 
and the proportionally weaker short-wavelength PAH bands in early-type 
galaxies. 

\subsection{Environments of Anomalous PAH Ratios}

We have used data from three significant studies of PAH emission from nearby galaxies 
to probe the relationship between PAH emission, AGN, and host galaxy morphology to 
further explore the different PAH band ratios seen in early-type galaxies. The largest 
study of early-type galaxies is the RSA \IRS{} Atlas
by \citet{Rampazzo13}, 
 and about half of their sample of ETGs exhibit PAH emission. \citet{Smith07} 
 performed a detailed analysis of PAH emission from galaxies in the
 \Spitzer{} Infrared 
Nearby Galaxy Survey \citep[SINGS:][]{Kennicutt03}. This sample spans a wide range of 
luminosity and infrared to visible wavelength flux ratio, and includes early and 
late-type galaxies. They have also classified the galaxies as either
``HII'' galaxies, 
which are dominated by star formation, LINERs, or Seyferts. Finally,
\citet{Diamond10} 
studied 35 Seyfert galaxies in the Revised Shapley-Ames catalog. All of these Seyferts 
have late-type host galaxies (later than S0), and generally exhibit weaker 
short-wavelength PAH emission compared to HII galaxies. This study included 21 
galaxies with PAH emission in spectra not centered on the nuclear region, 
and these off-nuclear spectra exhibit PAH ratios similar to HII galaxies. 

\begin{figure}
    \centering
    \includegraphics[width=\columnwidth]{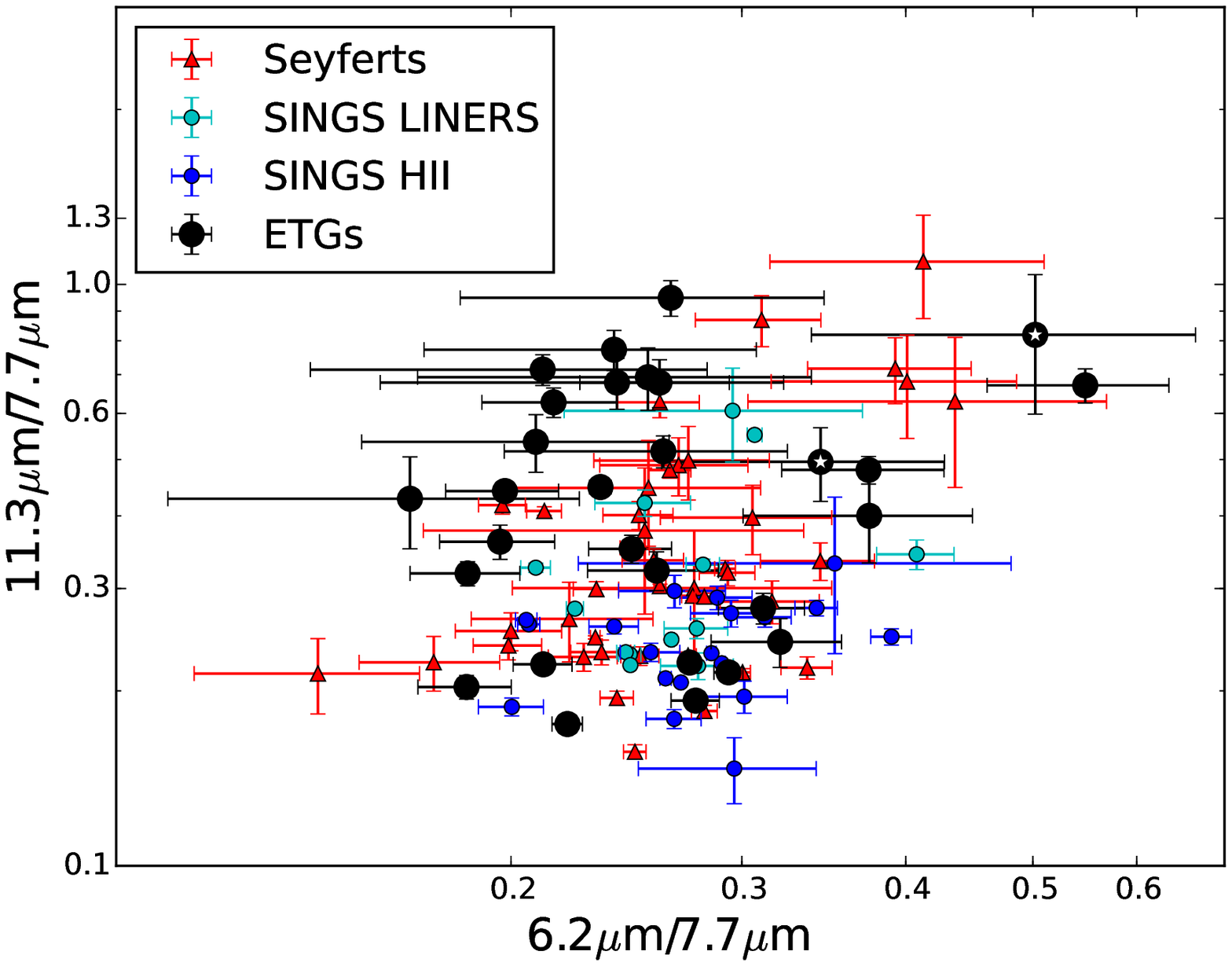}
    \caption{Plot of \ratioelevenseven{} vs. \ratiosixseven{} line
        ratios for different samples of galaxies. Seyfert line ratios 
        were taken from \citet{Diamond10}; SINGS ratios were from
        \citet{Smith07}; the ETG sample contains galaxies from
        \citet{Smith07} and \citet{Rampazzo13}. Stars indicate that the
        ETG is \Class{4} according to \citet{Rampazzo13}.}
    \label{fig:pah113-77-62}
\end{figure}

\citet{Diamond10} investigated if the weaker 6.2, 7.7, and \eightsix{} bands in 
Seyferts could be due to radiative or mechanical processing. Previous work 
\citep{Szczepanski93,Hudgins95,DeFrees93,Langhoff96} has shown that the 
$C-C$ stretching modes that give rise to the 6.2 and
\sevenseven{} features, and the 
$C-H$ in-plane bending modes that give rise to the \eightsix{} feature, are more 
readily produced in ionized PAHs. The ratio of these features to the $C-H$ 
out-of-plane bending mode that gives rise to the \eleventhree{} feature 
\citep{Duley81,Allamandola89} will be lower for more neutral PAHs. 
    \cref{fig:pah113-77-62} shows the ratio of
    \ratioelevenseven{} vs.\ \ratiosixseven{} for measurements from these 
    three studies. The early-type galaxies 
include both the \citet{Rampazzo13} sample and early-type galaxies in 
\citet{Smith07}, the Seyfert sample includes late-type Seyferts from both 
\citet{Smith07} and \citet{Diamond10}, and the LINERs and HII galaxies are only 
galaxies with late-type morphology from \citet{Smith07}. This diagram clearly shows 
that the early-type galaxies have larger \ratioelevenseven{} than the HII galaxies, 
but also that the ratio is larger than for the Seyferts and perhaps the 
LINER sample. Only galaxies with detections in both ratios are shown, as the 
literature sources generally do not quote upper limits for non-detections. 
Very different values of \ratioelevenseven{} vs.\ \ratiosixseven{} are 
expected for 
neutral vs.\ ionized PAHs. \citet{Allamandola99} and \citet{Draine01} showed 
that neutral PAHs lead 
to the intensity ratio \(\ratioelevenseven{} > 0.4\), whereas
\(\ratioelevenseven{} < 0.2\) is more characteristic of ionized PAHs, 
although the value of this ratio also 
depends on the PAH size distribution, as does the \ratiosixseven{} ratio. Only 
about half of the ETGs have \(\ratioelevenseven{} > 0.4\) as expected for 
neutral PAHs.

\begin{figure*}
    \centering
    \includegraphics[width=\textwidth]{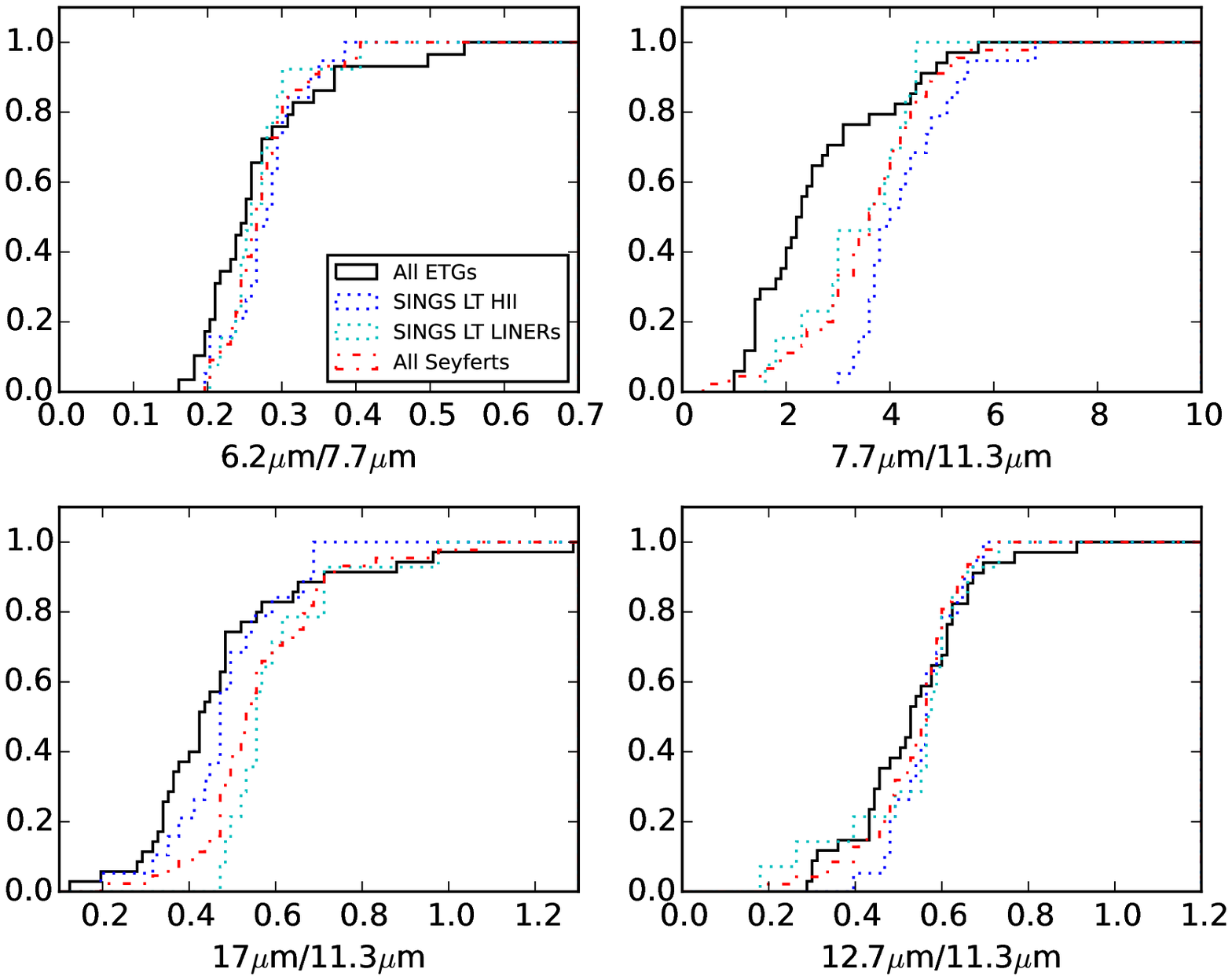}
    \caption{Cumulative distribution functions of the galaxy classes in
    \cref{fig:pah113-77-62} for \emph{(Top left)} \ratiosixseven{},
    \emph{(Top right)} \ratioseveneleven{}, \emph{(Bottom right)} 
    \ratiotwelveeleven{}, \emph{(Bottom left)} \ratioseventeeneleven{}.
  There is a statistically significant difference between the ETGs and
SINGS late-type (LT) HII galaxies, as well as the ETGs and Seyferts in
\ratioseveneleven{}. The distribution between ETGs and LINERs in
\ratioseveneleven{} was marginally significant. ETGs are also
significantly different from Seyferts and LINERs in
\ratioseventeeneleven{}, but are similar to HII galaxies. There are no 
significant differences between the galaxy distributions in any of the 
other line ratios.}
    \label{fig:pahcdf}
\end{figure*}

\cref{fig:pahcdf} plots the cumulative distribution functions of these four 
classes of galaxies. We used a Kolmogorov-Smirnov two-sample test to measure
the significance of the
differences. We found that the \ratioseveneleven{} difference between the 
early-type galaxies and the Seyferts, as well as 
between the early-type galaxies and the HII galaxies, is very significant 
($p \ll 0.01$). The difference between the early-type galaxies and LINERs is marginally 
significant ($p \sim 0.02$) as well. Note that the LINER sample is somewhat smaller 
than the other samples, and that the cumulative distributions contain all 
galaxies with detections in the single ratio, and thus may have more points than 
\cref{fig:pah113-77-62}. 

\begin{figure}
    \centering
    \includegraphics[width=\columnwidth]{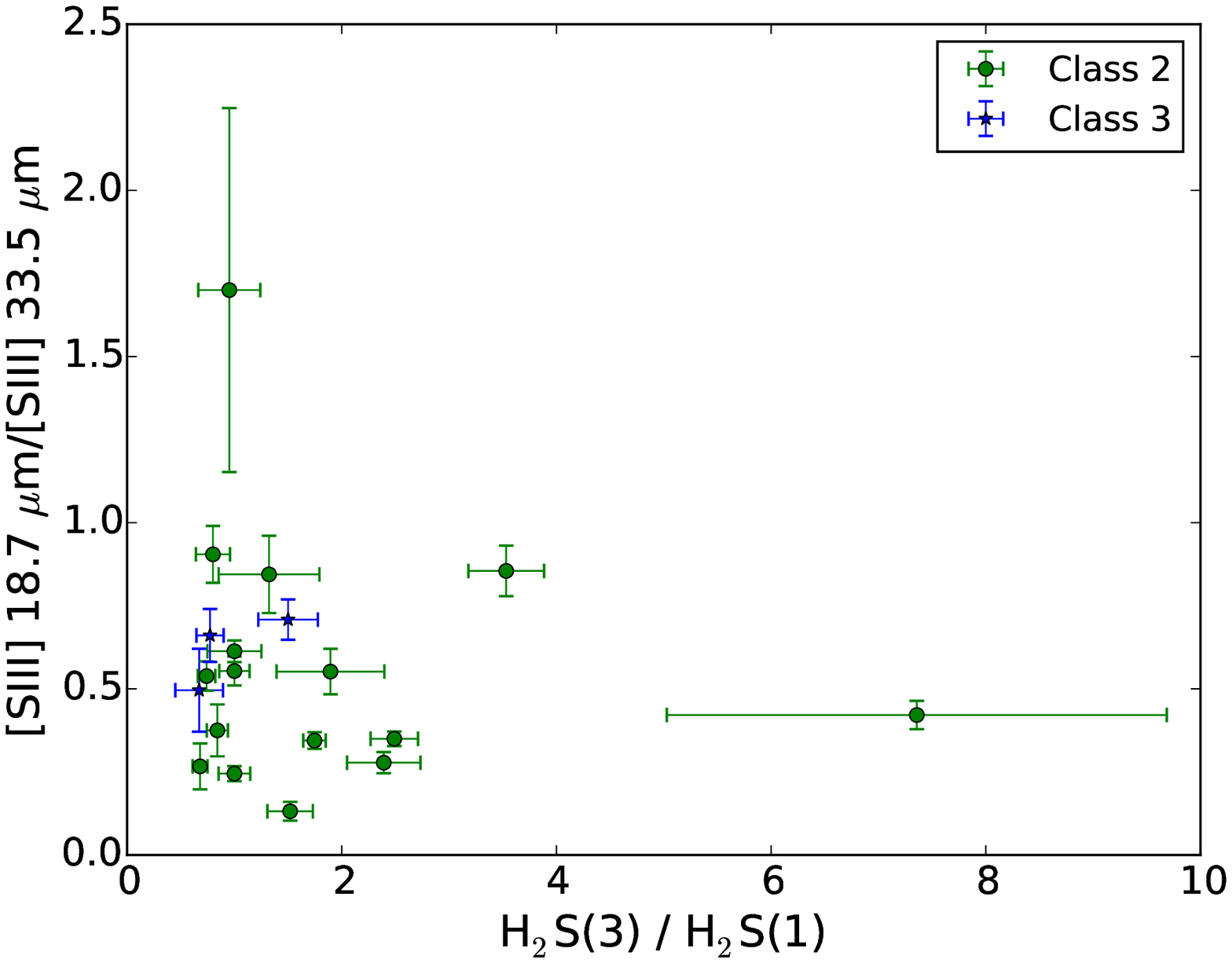}
    \caption{\Class{2} and \Class{3} sulfur and
    molecular hydrogen line ratios as measured in \citet{Rampazzo13}. We
    expect \ratiomolecular{} to be a proxy for the gas temperature of
    the emitting region, and \ratiosulfur{} to be a proxy for the
    electron density. Objects without detections in any of the lines
    have been excluded from the figure. There does not appear to be a 
    significant distinction between these two classes in this plane.}

    \label{fig:lineratios}
\end{figure}

\citet{Galliano08} 
noted that the typical ionization state of PAHs is set by \(G_0 T^{1/2}/n_e\) 
\citep{Bakes94}, where \(G_0\) is the intensity of the UV radiation
field, \(T\) is the gas temperature, and \(n_e\) is the electron density, and 
we investigated if the  distribution of \Class{2} and \Class{3} objects
followed trends in any of these physical quantities.
The gas temperature can be determined from the 
\ratiomolecular{} line ratio, and the electron 
density can be determined from \ratiosulfur{}.
These lines have been measured by \citet{Rampazzo13}, and any trends
should cause the two classes to separate in one of the dimensions. 
As seen in \cref{fig:lineratios}, \Class{2} and \Class{3} objects have 
similar values of these ratios,
so we conclude that trends with gas temperature or electron density are
not the origin of the difference between the
two classes. 

\begin{figure}
    \centering
    \includegraphics[width=\columnwidth]{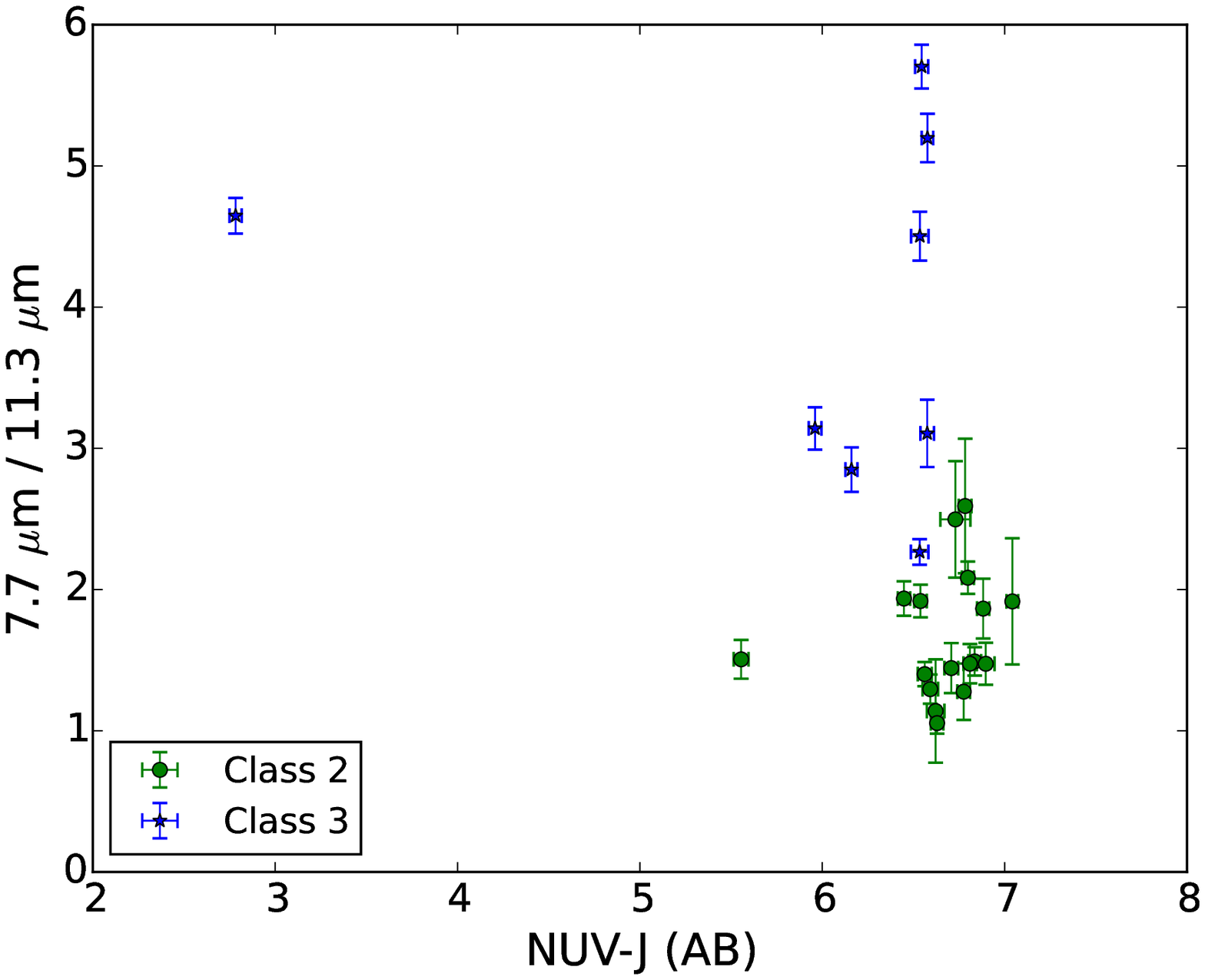}
    \caption{Distribution of \Class{2} and \Class{3} sources in NUV-J,
  which acts as a proxy for the slope of the ISRF, and
  \ratioseveneleven{}, which reflects the level of ``anomaly'' in the PAH ratios.} 
\label{fig:uvpahs}
\end{figure}

Since the temperatures and densities for the \Class{2} and \Class{3}
galaxies are similar, we expect the degree of ionization
for PAHs will depend on the shape of the interstellar radiation field
(ISRF) \citep{Weingartner01}. The ISRF in early and late-type galaxies varies
dramatically, both in normalization and shape. The dramatic lack of
young stars, and therefore UV emission, can even lead optical photons
to become the primary heating mechanism for interstellar dust
\citep{Groves12}. While we do not directly measure the intensity of the
ISRF, we
use the NUV-J color as a proxy for the slope. 
We show the trend in the slope of the radiation field with the
\ratioseveneleven{} ratio in \cref{fig:uvpahs}. 
Even though \Class{4} objects also have PAH emission, they were excluded
as likely AGN because their high 1.4 GHz and nuclear X-ray emission
may distort the ISRF \citep{Rampazzo13}. There does not appear 
to be any correlation between the shape of the ISRF and the 
\ratioseveneleven{} ratio, which corresponds directly with MIR class. 

While the relation between ionization state and UV intensity works well
for many environments \citep{Joblin96,Bregman05}, \citet{Diamond10} pointed 
out that many Seyferts have ratios consistent with completely neutral PAHs, 
or even more 
extreme ratios than the models allow, and in any case the more intense radiation 
field associated with AGN should not lead to less ionization. We find that the 
early-type galaxies often have even smaller values of the
\ratioseveneleven{} band ratio than the 
Seyferts.  \citet{Draine01} noted that smaller values of \ratioseveneleven{} 
could be produced in the charging conditions characteristic of the Cold 
Neutral Medium (CNM) or Warm Ionized Medium (WIM), which have lower densities 
and radiation field intensities than typical of the Photodissociation Regions 
(PDRs) associated with star formation. If most of the PAHs are associated with 
conditions more similar to the CNM 
or WIM, this could explain the dramatic difference in their ratios relative to 
star-forming galaxies, as well as why they are not similar to Seyferts, which 
typically also have PDRs.

\begin{figure}
    \centering
    \includegraphics[width=\columnwidth]{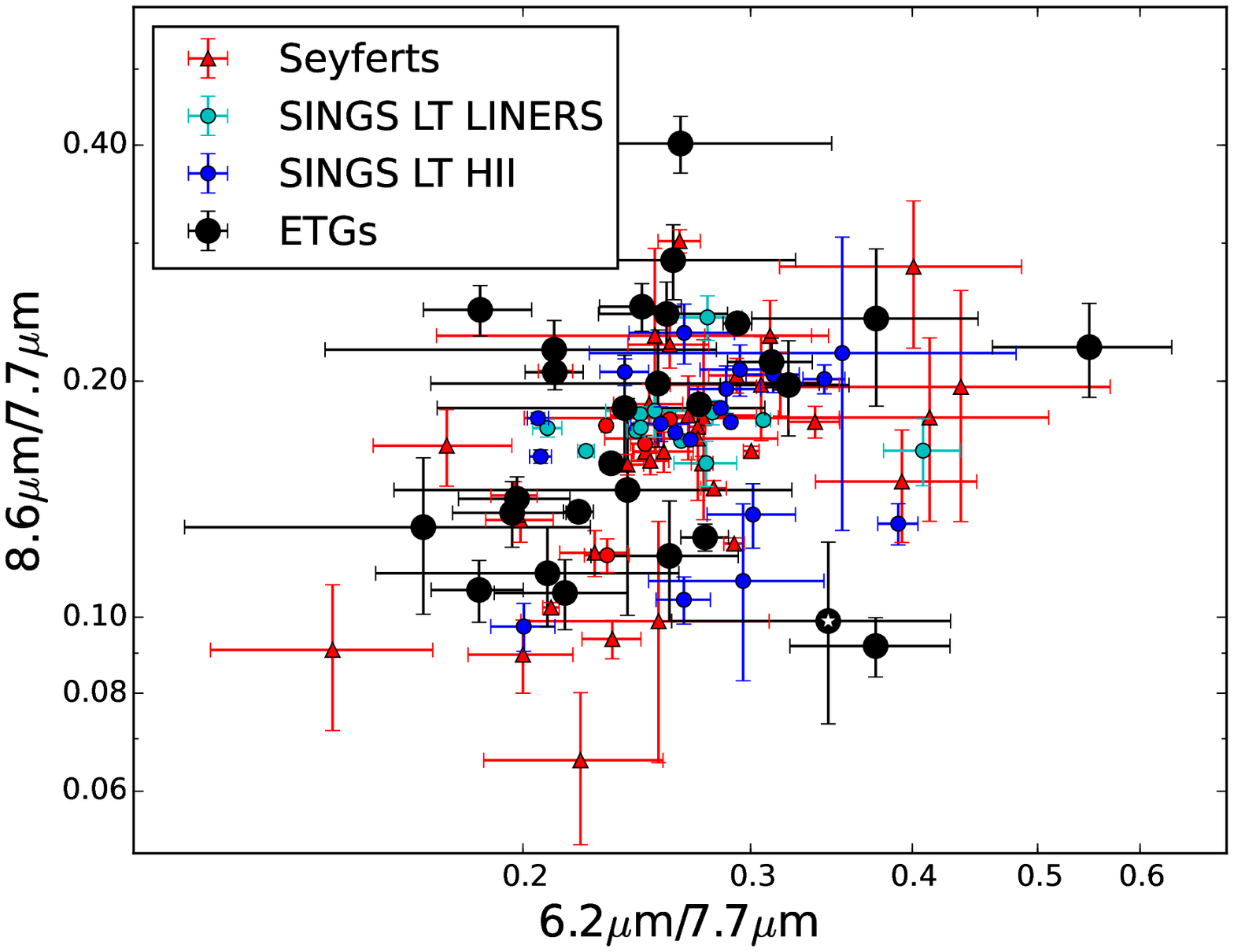}
    \caption{Plot of line ratios like \cref{fig:pah113-77-62} except for
        \ratioeightseven{} and \ratiosixseven{} line ratios. Larger
        grain size distributions tend to have suppressed
        \ratiosixseven{} emission. There are no statistically significant
      differences in the ratios between classes, indicating that a
    systematic difference in grain size distribution may not be
  correlated with anomalous PAH emission.}
    \label{fig:pah62-77-86}
\end{figure}

As the three shorter-wavelength PAH features are also associated with smaller 
PAHs, an alternative explanation of their proportionally lower fluxes is a 
different size distribution \citep{Schutte93,Draine01,Galliano08}.
Experimental work on PAHs has indicated that PAHs below 15-20 C atoms would be 
destroyed in most environments, PAHs smaller than 20-30 C atoms may be stripped of 
their H atoms, and PAHs of 30-50 C atoms would mostly be photoionized 
\citep{Jochims94,Allain96}.  If smaller PAHs have been destroyed, then this would also 
lead to lower \sixtwo{} relative to \sevenseven{} and lower
\sevenseven{} relative to 
\eightsix. \cref{fig:pah62-77-86} shows these two ratios for the same 
galaxies shown in \cref{fig:pah113-77-62}. Based on inspection of this figure, 
as well as Kolmogorov-Smirnov (KS) tests applied to the cumulative 
distributions, we do not see significant 
differences between the early-type galaxies and other populations. \citet{Diamond10} 
similarly did not find significant differences between Seyferts and either their 
off-nuclear positions or the HII galaxies from \citet{Smith07}. As illustrated in 
\citet{Draine01}, larger PAHs can also contribute to the larger
\ratioelevenseven{} ratios
seen in \cref{fig:pah113-77-62}, although they also have smaller 
\ratiosixseven{} ratios. Intriguingly, about half of the early-type galaxies have 
\(\ratiosixseven{} < 0.25\), although the difference is not formally significant. Careful 
estimates of upper limits on the weak \sixtwo{} feature in the several early-type 
galaxies with only \sevenseven{} detections, as well as more sensitive measurements with future 
facilities, would help to better characterize the smaller PAHs in these galaxies 
that are most sensitive to destruction processes. 

\begin{figure*}
    \centering
    \includegraphics[width=\textwidth]{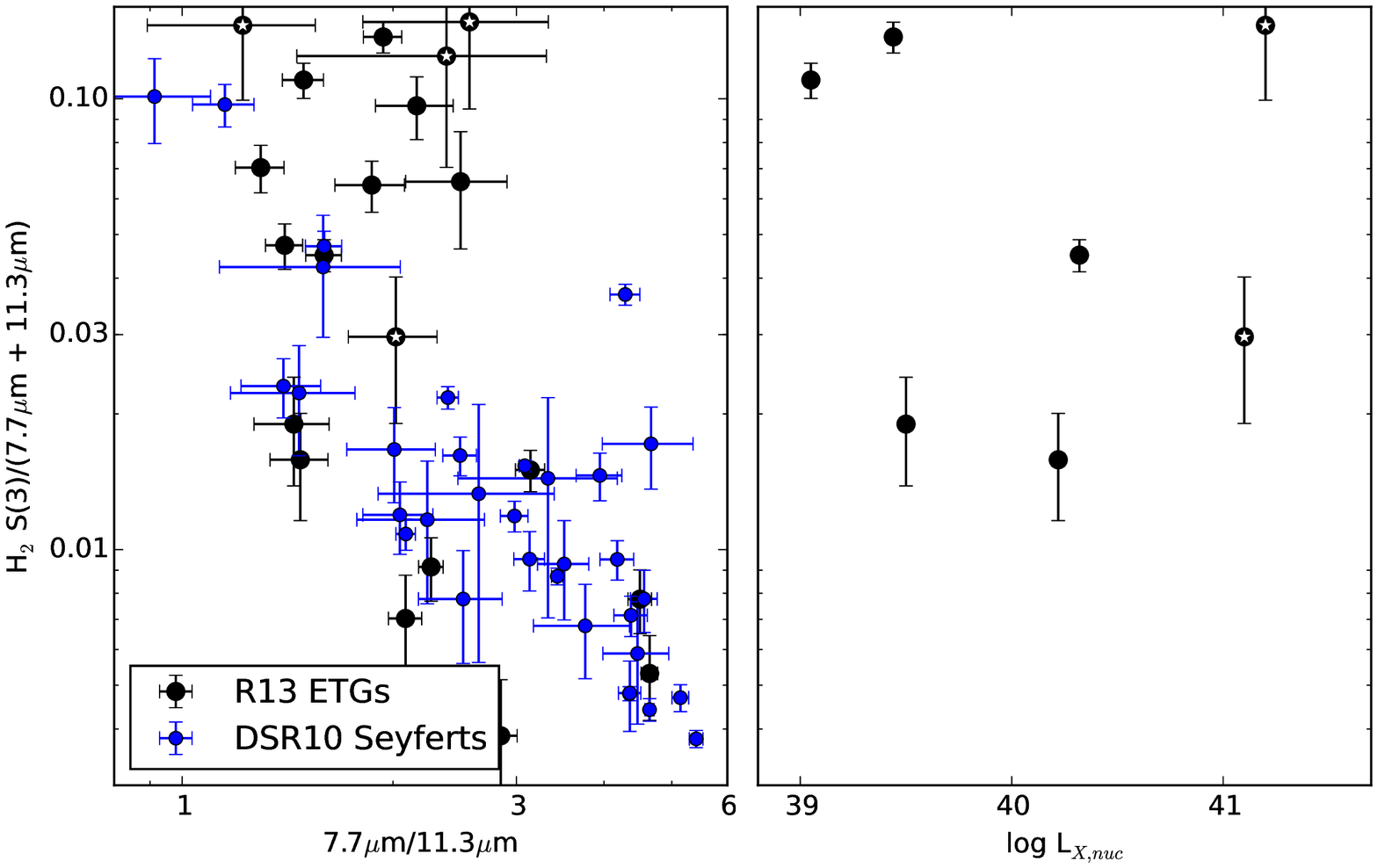}
    \caption{\textit{Left:} Relative strength of the
    \molecularthree{}
        transition to PAH emission, which traces shocks in cold gas,
        versus the \ratioseveneleven{} ratio. Black
        points are ETGs from \citet{Rampazzo13} while blue points are
        Seyferts from \citet{Diamond10}. Stars indicate \Class{4}
        galaxies. Both the ETGs and Seyferts indicate a correlation
        between the shocks and anomalous PAH ratios. \textit{Right:}
        Molecular hydrogen strength vs.\ nuclear X-ray luminosity for
        \Rampazzo{} galaxies with nuclear X-ray detections. As noted in
        \citet{Diamond10}, there does not appear to be a trend between
      shock strength and X-ray luminosity.}
    \label{fig:pahh2}
\end{figure*}

Another explanation proposed for the variations in
\ratioseveneleven{} is interstellar shocks. 
Shock velocities of on order 100 \kms\ can reduce the relative number of small grains 
\citep{Jones94,Micelotta10}, and thus lead to weaker emission in all three of the 
shorter-wavelength features. If shocks are present, then the galaxies may also exhibit 
strong \htwo{} emission. Many studies have found a correlation between \htwo{}
emission and the \ratioseveneleven{} ratio
\citep{Roussel07,Ogle07,Kaneda08,Vega10}. 
\cref{fig:pahh2} \emph{(left)} illustrates the correlation shown by 
\citet{Diamond10} for the RSA Seyferts, along with the early-type galaxies from 
\citet{Rampazzo13}. The early-type galaxies mostly follow the same trend as the 
Seyferts, although they generally have more \htwo{} emission and there is more 
dispersion in \ratioseveneleven{} for the early-type galaxies with the proportionally 
strongest \htwo{} emission. \citet{Diamond10} and \citet{Rampazzo13} 
found that AGN power did not correlate with this ratio, and
\cref{fig:pahh2} \emph{(right)} confirms this point with a plot of the ratio of 
\lineratio{\molecularthree}{(\sevenseven + \eleventhree)} vs.\ nuclear X-ray luminosity from \citet{Pellegrini10}. 
Processing by shocks appears to be a more viable contributor to the
\ratioseveneleven{} ratios in early-type galaxies, as it is for Seyferts. 

\begin{figure}
  \centering
  \includegraphics[width=\columnwidth]{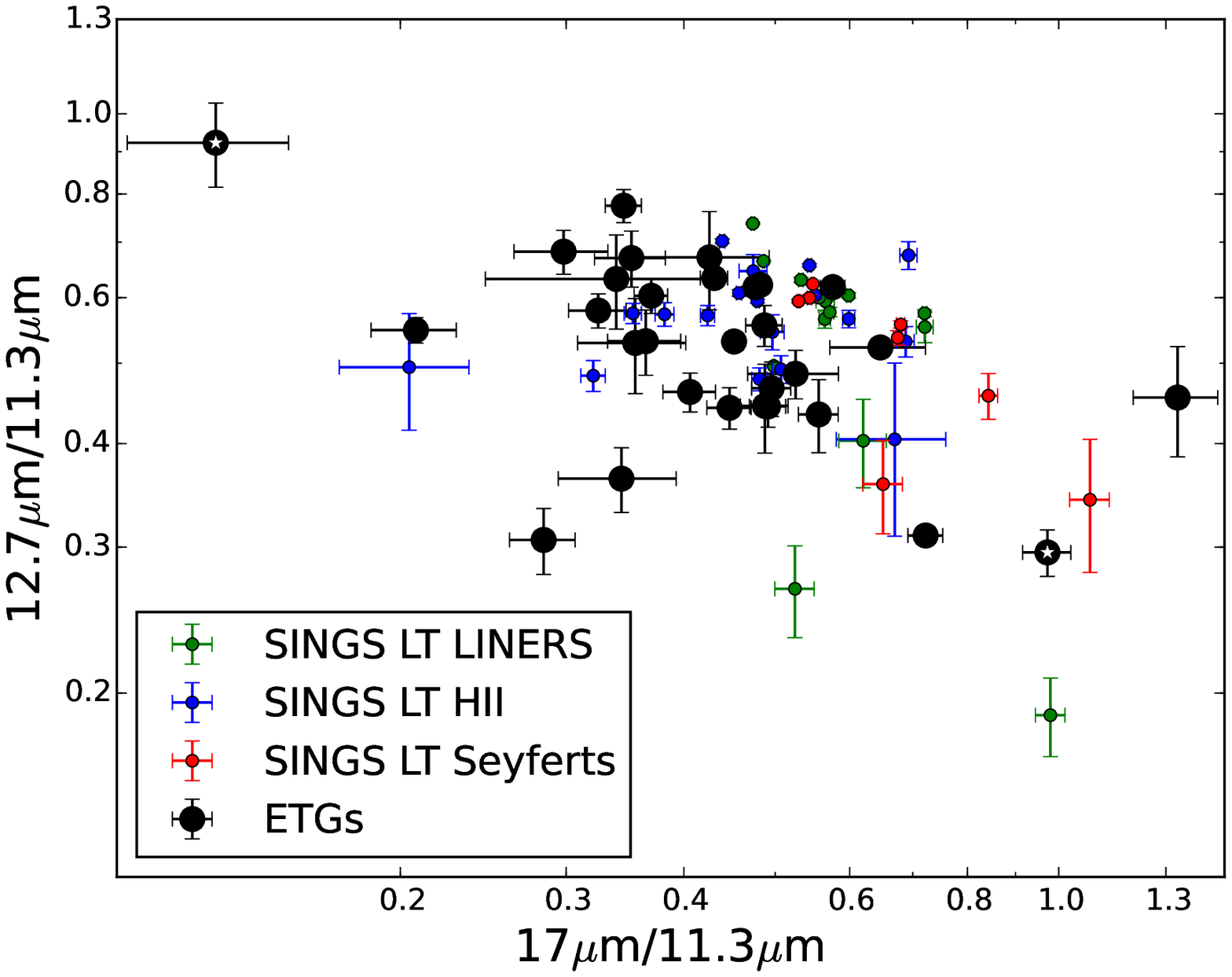}
    \caption{Plot like \cref{fig:pah113-77-62}, but with
    \ratiotwelveeleven{}, which arises from C---H multiplets, vs.
    \ratioseventeeneleven{}, which arises from single C---H bonds. While
  there does seem to be a distinction between HII galaxies and Seyferts
  in the \ratiotwelveeleven{} ratio, the ETG population is not statistically 
  distinct from either of them. For the \ratioseventeeneleven{} ratio,
ETGs are similar to HII galaxies, but distinct from LINERs and Seyferts.}
  \label{fig:pah127-113-17}
  \end{figure}

In addition to the preferential destruction of small PAHs, shocks could also affect the 
chemistry of the PAHs, even those that are fully hydrogenated. For example, the 
\eleventhree{} band is produced by single C---H bonds, while the \twelveseven{} feature is 
produced by C---H multiplets \citep{Hony01}. \citet{Diamond10} found that Seyferts 
exhibit significantly smaller ratios of \ratiotwelveeleven{} than off-nuclear and HII galaxies, 
which could be due to different processing. \cref{fig:pah127-113-17,fig:pahcdf} 
shows that the early-type galaxies are not significantly different 
from the Seyferts or the HII galaxies, so we see no evidence that this is an 
important physical difference in the early-type galaxy population. 

The \ratiotwelveeleven{} band ratio shown in \cref{fig:pah127-113-17} is 
plotted vs.\ the \ratioseventeeneleven{} band ratio. Both the \eleventhree{} 
band and the \seventeen{} band originate from neutral PAHs, and so their 
relative strength is primarily sensitive to the size distribution. 
\citet{Smith07} found that larger \ratioseventeeneleven{} intensity 
ratio correlates with metallicity, and suggest that it may be easier to form the 
larger grains that contribute to the \seventeen{} band in higher
metallicity environments. 
They find that \ratioseventeeneleven{} is largest in Seyferts and suggest that these high ratios 
are due to a combination of the higher metallicities of the Seyfert hosts and the 
relative destruction of the carriers of the shorter wavelength \eleventhree{} feature. 
As shown in \cref{fig:pah127-113-17}, we find that early-type galaxies 
have significantly \textit{lower} \ratioseventeeneleven{} ratios than the Seyferts (KS $p<0.0001$) and 
LINERs (KS $p<0.0001$) from \citet{Smith07}, and are similar to HII galaxies 
(see also \cref{fig:pahcdf}). Early-type galaxies may have lower
\ratioseventeeneleven{}
band ratios because they lack the hard radiation field of Seyferts. Alternatively, 
this ratio could be lower in some early-type galaxies because their PAHs have been 
externally accreted from lower-metallicity dwarf galaxies. 

We cannot separate the effect of metallicity on the
\ratioseventeeneleven{} ratio because of the sparse overlap between the
\ATLAS{} and \Rampazzo{} samples. Only 13 galaxies in our sample have
both \ATLAS{} metallicity and \ratioseventeeneleven{} measurements;
their Pearson correlation coefficient is -0.04, which indicates a very weak
anticorrelation. 

\subsection{Photometric PAH classification}

Currently the only way to discriminate between MIR classes is through
infrared spectra from targeted spectroscopic observations in space.
This is a significant investment for space-based missions which are
already highly oversubscribed. Now that the \WISE{} mission has 
imaged the entire sky in the MIR, a successful method of classifying
galaxies into different MIR classes based on \WISE{} data
would allow demographic analysis of anomalous PAH emission.

\begin{figure*}
    \includegraphics[width=\textwidth]{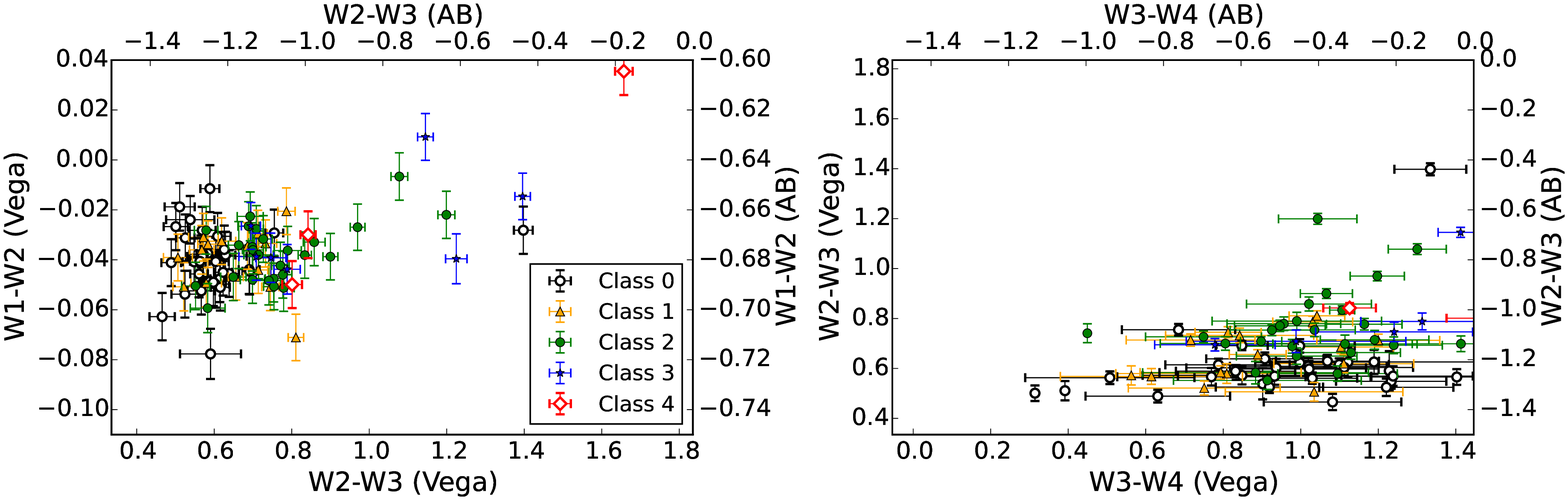}
    \caption{\emph{Left:} W1-W2 vs. W2-W3 for the sample of galaxies in
        \citet{Rampazzo13}. W1-W2 is sensitive to the stellar
        population, so we do not see separation based on properties of the
        interstellar dust. W2-W3 measures the presence of PAH
        features compared to the stellar population. This causes the \Class{0}
        objects separate from the \Class{2-4} objects. Galaxies located off the plot are: IC 5063,
        NGC 1052, NGC 1275, NGC 2685, NGC 4383, and NGC 5128.
        \emph{Right:} W2-W3 vs W3-W4 for the sample of galaxies in
        \citet{Rampazzo13}. There does not appear to be strong discrimination
        in W3-W4 due to the hot dust population, as seen in W2-W3. Located off the plot to the top and to the
        right are: all \Class{4} objects, NGC 2685, NGC 3245, NGC 
        4383, and NGC 4435.}
    \label{fig:dual_mir_plot}
\end{figure*}

\begin{figure}
    \includegraphics[width=\columnwidth]{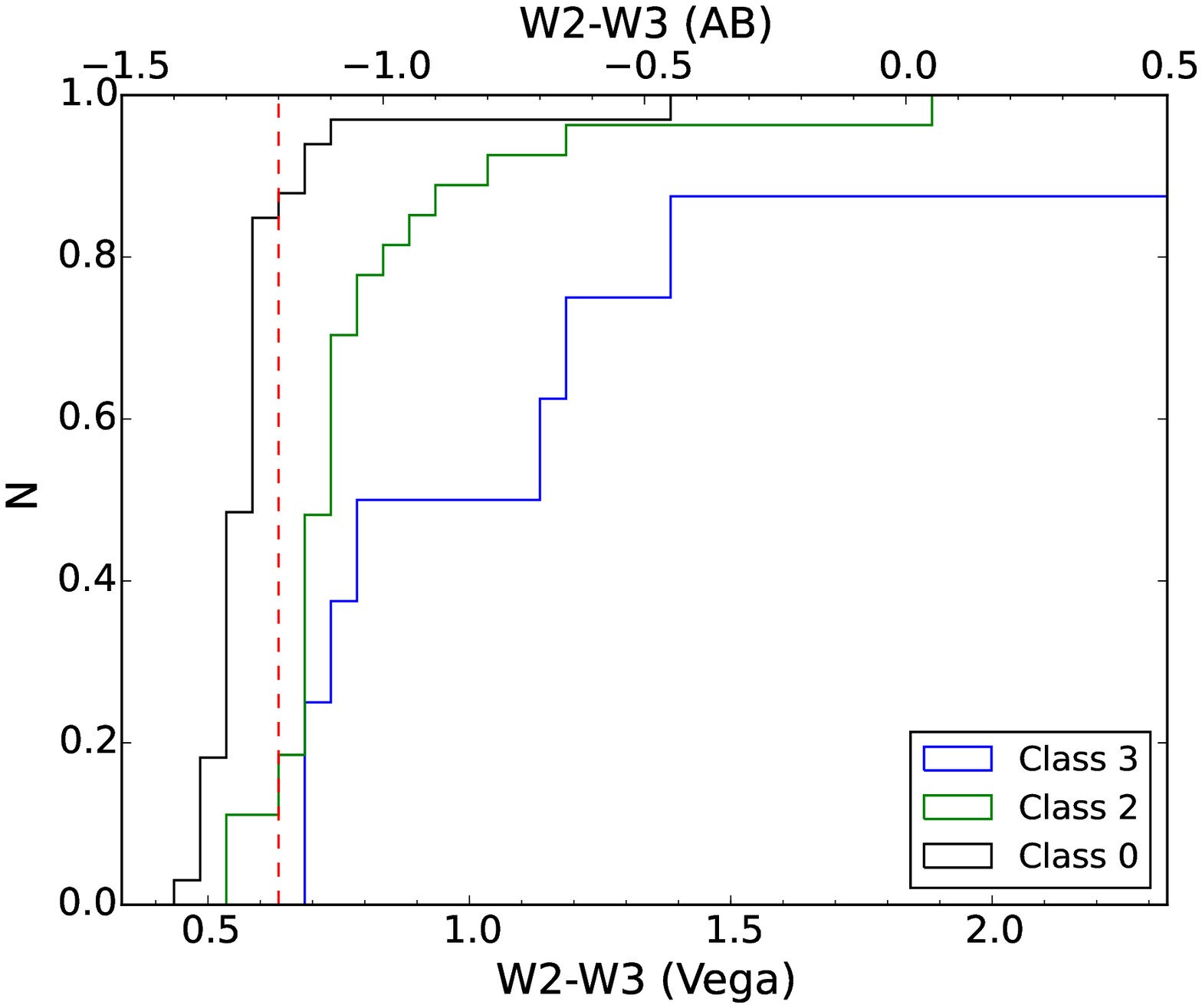}
    \caption{Cumulative histogram of the W2-W3 colors of the \Class{0},
    \Class{2}, and \Class{3} objects. The \Class{0} objects are from a
    population statistically distinct from both the \Class{2} and
    \Class{3} objects. We propose a color-cut to distinguish between 
    \Class{0} and \Class{2--3} objects at W2-W3=-1.2 (AB), which is
    denoted by the red dashed line. \Class{4} objects 
    are from a population statistically distinct from all other classes. 
    There are no other separations which can be made using W2-W3 colors.}
    \label{fig:class023_cumhist}
\end{figure}

Because most of the PAH emission features lie in the very broad W3
filter, we expect most of the discriminating power to occur with the
W2-W3 and W3-W4 colors.
We plot the \Rampazzo{} sample in \cref{fig:dual_mir_plot}. Many of the
\Class{4} objects lie far away from the bulk of the distribution, likely 
due to the hot dust continuum in their spectra. 
There also seems to be good discrimination between \Class{0} and
\Class{2/3} objects in W2-W3. The \Class{0} objects are
systematically bluer than both the \Class{2} (K-S \(p < 10^{-4}\)) and
\Class{3} (K-S \(p < 10^{-6}\)) objects. A cumulative histogram of both classes
in W2-W3 is shown in \cref{fig:class023_cumhist}. 85\% of \Class{0},
11\% of \Class{2}, and no \Class{3} objects are bluer than 
\(W2-W3= 0.635\) (Vega, or -1.2 AB), providing a convenient color-cut 
between dusty and passive ETGs. No other major 
distinctions between classes are statistically significant.

One reason why \WISE{} is unable to discriminate between classes very well
is because the bands are not well-placed with respect to PAH features. 
The only band sensitive to PAH emission is W3, which covers all of the 
\sevenseven, \eightsix, \eleventhree, and  \twelveseven{} PAH complexes
\citep{Wright10}. The W3 band accomplishes this amount of spectral
coverage with a wide effective bandwidth of \mbox{5.5 \(\mu\)m}
\citep{Jarrett11}. By comparison, the typical equivalent width of a
line is around \mbox{0.1 \(\mu\)m} \citep{Rampazzo13}. Therefore, the difference 
between the presence and
absence of all five PAH lines is around 0.1 mag, which is similar to the
photometric error of our measurements. While this should suffice to
separate \Class{0} objects from \Class{2/3} objects, separation of
\Class{2} and \Class{3} objects requires more precise photometry.
Another complication 
is the large intrinsic scatter in \Class{2} and \Class{3}, which we
attribute to varying levels of contribution by the dust continuum.

Despite the fact that \WISE{} colors on their own cannot discriminate
between the MIR classes, they can be useful for informing follow-up
targets. This will be particularly true with the Mid-Infrared
Instrument (MIRI) on the \textit{James Webb Space Telescope (JWST)}. For 
galaxies which have W2-W3 colors consistent with some interstellar dust, 
better demographics on the prevalence of anomalous PAH ratios can be 
determined from follow-up with the MIRI Medium Resolution Spectrograph (MRS).

\section{Conclusions}
\label{sec:conclusion}

We performed \WISE{} and \GALEX{} aperture photometry on a combined sample
of 91 ETGs from \Rampazzo{} which had uniform MIR line flux
measurements, and 260 ETGs from the \ATLAS{} survey, which have
significant ancillary data. The photometric accuracy is
about 0.05 mag in the MIR and 0.1 mag in the UV\@. We chose an identical aperture
to one used by the \TM{} XSC.

The MIR is a useful wavelength region to calculate stellar masses since the 
effects of extinction and young stars are reduced compared to shorter
wavelengths. We converted the \ATLAS{} \( \left( M/L \right)_{stars}\) from 
r-band to W1 and corrected for the change in aperture between the two studies. 
We found that the \ATLAS{} \(M/L_{W1}\) measurements are significantly 
larger than those predicted by SPS models. We did not find a trend
between \(M/L_{W1}\) and W1-W2 for old stellar populations predicted by
SPS models. We obtained an average \(\log M/L_{W1} = 0.07 \pm 0.13\), which is 
based on dynamical \(M/L_{W1}\) rather than fits to SPS models. Our 
color selection of dustless galaxies biases our value toward higher
\(M/L_{W1}\); but without the color-cut, we still observe a \(2\sigma\) 
discrepancy.

Our high \(M/L_{W1}\) compared to SPS models agrees with
recent research indicating that the IMF varies, and is more 
Salpeter-like at high velocity dispersions. When the \(M/L_{W1}\) from our 
galaxies is compared to 
FSPS models with \Chabrier{} and \Salpeter{} IMFs, and we find that the 
dynamically-derived masses mostly fall in-between the two IMFs, mirroring 
the effect seen from \(M/L_{W1}\) derived from optical spectra. 

We clearly identify circumstellar dust in the W3 and W4 bands for many ETGs. 
Surprisingly, this circumstellar dust emission appears to be ubiquitous at 
much greater ages than predicted by SPS models, even after accounting for 
potential contamination in our dustless sample. We also find that W4 emission 
is underpredicted by models by about a factor of 2.5. This underprediction
is also seen in other SPS models.

With all-sky coverage, the \WISE{} mission makes stellar mass and SFR 
estimates very accessible. The \WISE{} W3 and W4 bands are sensitive
to dust warmed by recent star formation. For objects with low sSFR, 
circumstellar dust emission can contaminate these tracers, leading to an 
inflated result.  We determined that circumstellar dust contributes 
approximately \( 2 \times 10^{-12}\) yr\(^{-1}\) to the sSFR signal in both W3 
and W4 and we provide relations to correct for the circumstellar dust 
contribution to W3 and W4.

Lastly, we compared MIR spectra for spirals, LINERs, Seyferts, and ETGs
in order to investigate the source of anomalous PAH ratios. We found that
ETGs and Seyferts have significantly distinct
\ratioelevenseven{} ratios from spirals, as well as from each other.
This strongly suggested that the cause of anomalous PAH ratios is more
than just the presence of an AGN\@. The ETGs also have
significantly distinct \ratioseventeeneleven{} values from the 
LINERs and Seyferts.

Using available \ratiosulfur{} and \ratiomolecular{} ratios, we noted
that gas temperature and electron density do not correlate with the
presence of anomalous PAH ratios. The incident UV flux on the PAHs also
does not appear to correlate with \ratioseveneleven{} ratio. Based on
these results, we conclude that the anomalous PAH ratios are
unlikely to be due to global variation in the ISRF\@. 

We did not see evidence for the preferential destruction of smaller
PAHs based on the \ratiosixseven{} and \ratioeightseven{} ratios. However,
due to the nonuniform availability of upper limits in published line
intensities, we excluded galaxies without detections. A uniform treatment of 
upper limits may reveal more information
about small PAHs in these galaxies. We did not find evidence of 
dehydrogenation of small PAHs in the \ratiotwelveeleven{} ratio. 

The strongest correlation with anomalous PAH emission is the strength of
shocks in \htwo. We found that both ETGs and Seyferts follow this trend, 
although ETGs generally have more \htwo{} emission. Exactly how shocks bring about the anomalous emission is 
still largely uncertain. However, we determined from X-ray luminosities
that it is likely unrelated to the activity of the central supermassive
black hole. 

One particularly interesting result we found was that the ETGs have
 \ratioseventeeneleven{} ratios significantly lower than Seyferts and LINERS, 
 but similar to spirals. This ratio is sensitive to the presence of
 large grains, and has been observed to trend with metallicity. Our
 result can be explained by the lack of the hard radiation environment
 found in Seyferts, or if the PAHs were accreted from an external,
 low-metallicity galaxy.

Finally, we attempted to use \WISE{} to classify the PAH properties of
ETGs. We found that \WISE{} colors are
only able to determine the presence of dust in ETGs, but not the MIR
class. We conclude that finer distinctions are not possible
because of the spectral placement of the \WISE{} filters and our
photometric precision. However, the W2-W3 colors may be able to inform detailed 
follow-up spectra of early-type galaxies in order to more efficiently calculate
the prevalence of anomalous PAH ratios.

\section*{Acknowledgments}
GVS and PM would like to thank C. Conroy for his
valuable insights into SPS models, A. Gil de Paz for helpful information
about \GALEX{} photometry, and J.D. Smith for useful discussions about
PAHs, and the anonymous referee. GVS would also like to thank T. Jarrett 
for providing additional details regarding \WISE{} fluxes, luminosities, masses, 
and SFRs for nearby galaxies. This publication makes use of data products from 
the \textit{Wide-field Infrared Survey Explorer}, which is a joint project of 
the University of California, Los Angeles, and the Jet Propulsion 
Laboratory/California Institute of Technology, funded by the National 
Aeronautics and Space Administration. Some/all of the data presented in this 
paper were obtained from the Mikulski Archive for Space Telescopes (MAST). 
STScI is operated by the Association of Universities for Research in Astronomy, 
Inc., under NASA contract NAS5-26555. Support for MAST for
non-\textit{HST} data is provided by the NASA Office of Space Science via 
grant NNX13AC07G and by other 
grants and contracts. This research has made use of the NASA/IPAC Extragalactic
Database (NED) which is operated by the Jet Propulsion Laboratory,
California Institute of Technology, under contract with the National
Aeronautics and Space Administration. 

This research made use of Astropy, a community-developed core Python package 
for Astronomy \citepalias{Astropy}. This research made use of APLpy, an 
open-source plotting package for Python hosted at http://aplpy.github.com. This
research made use of the IPython package \citep{IPython}. IRAF is
distributed by the National Optical Astronomy Observatory, which is
operated by the Association of Universities for Research in Astronomy
(AURA) under a cooperative agreement with the National Science
Foundation. This research made use of matplotlib, a Python library for
publication quality graphics \citep{Matplotlib}. PyRAF is a product of
the Space Telescope Science Institute, which is operated by AURA for
NASA\@. This research made use of SciPy \citep{Scipy}.


\bibliographystyle{mnras}
\bibliography{references}

\end{document}

%% file: tables/params.tex
\begin{table*}

\caption{Aperture photometry parameters for galaxies in the \ATLAS{}
    and \Rampazzo{} samples
    \label{tab:params}}
\begin{tabular}{cccccccccccc}
\hline
Galaxy & Morph & D & S & \(a_{W1}\) & \(a_{W4}\) & \(b/a_{W1}\) & \(b/a_{W4}\) & PA & Survey & NUV Tile & FUV Tile \\
 (1)  & (2)  & (3)  & (4)  & (5)  & (6)  & (7)  & (8)  & (9)  & (10)  & (11)  & (12)   \\
\hline
IC0560 & S0-a & 27.2 & A & 31.1 & 35.23 & 0.47 & 0.56 & 10 & AllWISE & MISDR1\_24342\_0266 & MISDR1\_24342\_0266 \\
IC0598 & S0-a & 35.3 & A & 31.4 & 37.39 & 0.43 & 0.55 & 10 & AllWISE & -- & -- \\
IC0676 & S0-a & 24.6 & A & 37.9 & 40.93 & 0.62 & 0.62 & -20 & AllWISE & GI4\_042019\_J111315p093030 & GI4\_042019\_J111315p093030 \\
IC0719 & S0 & 29.4 & A & 42.73 & 48.45 & 0.35 & 0.43 & 50 & AllWISE & AIS\_232\_sg80 & AIS\_232\_sg80 \\
IC0782 & SBab & 36.3 & A & 18.53 & 22.42 & 0.65 & 0.75 & 60 & AllWISE & GI6\_001033\_GUVICS033 & GI3\_079021\_NGC4261 \\
IC1024 & S0 & 24.2 & A & 33.1 & 38.67 & 0.4 & 0.51 & 30 & AllWISE & MISDR1\_33738\_0535 & MISDR1\_33738\_0535 \\
IC1459 & E1 & 29.0 & R & 97.03 & 100.97 & 0.84 & 0.86 & 40 & AllWISE & GI1\_093001\_IC1459 & GI1\_093001\_IC1459 \\
IC2006 & E1 & 20.0 & R & 46.86 & 48.75 & 0.88 & 0.89 & 50 & AllWISE & AIS\_423\_sg76 & AIS\_423\_sg76 \\
IC3370 & E2 pec & 27.0 & R & 66.94 & 69.65 & 0.85 & 0.87 & 55 & AllWISE & -- & -- \\
IC3631 & S0-a & 42.0 & A & 24.4 & 28.31 & 0.6 & 0.7 & 90 & AllWISE & GI6\_001074\_GUVICS074 & AIS\_223\_sg22 \\
\hline
\end{tabular}

    \emph{Notes:} Galaxy information and aperture photometry
    parameters. Columns: (1)~galaxy name; (2)~morphological type; (3)~distance
    in Mpc; (4)~Originating Sample (A:~\ATLAS{} R:~\Rampazzo{} B:~Both); (5)~W1
    aperture semimajor axis in arcseconds; (6)~W4 aperture semimajor axis in
    arcseconds; (7)~W1 aperture axis ratio; (8)~W4 aperture axis ratio; 
    (9)~Position Angle (10)~\WISE{} data release catalog; (11)~\GALEX{} NUV 
    tilename; (12)~\GALEX{} FUV tilename. References for morphological type and
    distance come from \citet{Cappellari11} and \citet{Rampazzo13}. The full
    version of this table with the full galaxy sample is included in the online
    journal article.
\end{table*}

%% file: tables/mags.tex
\begin{table*}

\caption{Galaxy magnitudes for the \ATLAS{} and \Rampazzo{} samples.
    \label{tab:magtable}}
\begin{tabular}{ccccccccccccc}
\hline
Galaxy & FUV & \(\sigma_{FUV}\) & NUV & \(\sigma_{NUV}\) & W1 & \(\sigma_{W1}\) & W2 & \(\sigma_{W2}\) & W3 & \(\sigma_{W3}\) & W4 & \(\sigma_{W4}\) \\
 (1)  & (2)  & (3)  & (4)  & (5)  & (6)  & (7)  & (8)  & (9)  & (10)  & (11)  & (12)  & (13)   \\
\hline
IC0560 & 21.49 & 0.117 & 19.40 & 0.035 & 13.27 & 0.006 & 13.93 & 0.009 & 14.35 & 0.056 & 13.67 & 0.154 \\
IC0598 & -- & -- & -- & -- & 12.93 & 0.006 & 13.59 & 0.008 & 14.59 & 0.062 & 14.35 & 0.296 \\
IC0676 & 18.11 & 0.054 & 16.92 & 0.031 & 12.46 & 0.006 & 13.01 & 0.008 & 11.06 & 0.016 & 9.681 & 0.013 \\
IC0719 & 17.29 & 0.072 & 16.68 & 0.037 & 12.34 & 0.006 & 12.96 & 0.008 & 12.07 & 0.017 & 11.99 & 0.045 \\
IC0782 & 21.57 & 0.108 & 19.21 & 0.035 & 13.93 & 0.007 & 14.62 & 0.012 & 15.78 & 0.142 & $>$14.8 & -- \\
IC1024 & 17.31 & 0.052 & 16.77 & 0.031 & 12.60 & 0.006 & 13.04 & 0.007 & 11.02 & 0.015 & 10.29 & 0.013 \\
IC1459 & 16.74 & 0.051 & 15.64 & 0.030 & 9.566 & 0.006 & 10.25 & 0.007 & 11.30 & 0.022 & 11.58 & 0.095 \\
IC2006 & 18.47 & 0.109 & 17.14 & 0.041 & 11.25 & 0.006 & 11.93 & 0.007 & 13.26 & 0.037 & 13.67 & 0.226 \\
IC3370 & -- & -- & -- & -- & 10.62 & 0.006 & 11.26 & 0.007 & 12.00 & 0.020 & 12.15 & 0.072 \\
IC3631 & 17.61 & 0.078 & 16.87 & 0.030 & 13.98 & 0.007 & 14.63 & 0.012 & 15.39 & 0.180 & $>$14.4 & -- \\
\hline
\end{tabular}

    \emph{Notes:} Calculated magnitudes for galaxies in our 
    sample reported in the AB system. They have not been corrected for 
    extinction. Columns: (1)~Galaxy name; (2)~FUV magnitude; (3)~FUV magnitude
    uncertainty; (4)~NUV magnitude; (5)~NUV magnitude uncertainty; (6)~W1
    magnitude; (7)~W1 magnitude uncertainty; (8)~W2 magnitude; (9)~W2 magnitude
    uncertainty; (10)~W3 magnitude; (11)~W3 magnitude uncertainty; (12)~W4
    magnitude; (13)~W4 magnitude uncertainty. Uncertainties contain statistical
    and calibration uncertainties. All upper limits are \(3\sigma\) upper
    limits. Blank UV entries mean that no \GALEX{} observations were taken. 
    The full version of this table with the full galaxy sample is included
    in the online journal article
\end{table*}

%% file: main.bbl
\begin{thebibliography}{}
\makeatletter
\relax
\def\mn@urlcharsother{\let\do\@makeother \do\$\do\&\do\#\do\^\do\_\do\%\do\~}
\def\mn@doi{\begingroup\mn@urlcharsother \@ifnextchar [ {\mn@doi@}
  {\mn@doi@[]}}
\def\mn@doi@[#1]#2{\def\@tempa{#1}\ifx\@tempa\@empty \href
  {http://dx.doi.org/#2} {doi:#2}\else \href {http://dx.doi.org/#2} {#1}\fi
  \endgroup}
\def\mn@eprint#1#2{\mn@eprint@#1:#2::\@nil}
\def\mn@eprint@arXiv#1{\href {http://arxiv.org/abs/#1} {{\tt arXiv:#1}}}
\def\mn@eprint@dblp#1{\href {http://dblp.uni-trier.de/rec/bibtex/#1.xml}
  {dblp:#1}}
\def\mn@eprint@#1:#2:#3:#4\@nil{\def\@tempa {#1}\def\@tempb {#2}\def\@tempc
  {#3}\ifx \@tempc \@empty \let \@tempc \@tempb \let \@tempb \@tempa \fi \ifx
  \@tempb \@empty \def\@tempb {arXiv}\fi \@ifundefined
  {mn@eprint@\@tempb}{\@tempb:\@tempc}{\expandafter \expandafter \csname
  mn@eprint@\@tempb\endcsname \expandafter{\@tempc}}}

\bibitem[\protect\citeauthoryear{{Abazajian} et~al.,}{{Abazajian}
  et~al.}{2009}]{Abazajian09}
{Abazajian} K.~N.,  et~al., 2009, \mn@doi [\apjs]
  {10.1088/0067-0049/182/2/543}, \href
  {http://adsabs.harvard.edu/abs/2009ApJS..182..543A} {182, 543}

\bibitem[\protect\citeauthoryear{{Alatalo} et~al.,}{{Alatalo}
  et~al.}{2013}]{Alatalo13}
{Alatalo} K.,  et~al., 2013, \mn@doi [\mnras] {10.1093/mnras/sts299}, \href
  {http://adsabs.harvard.edu/abs/2013MNRAS.432.1796A} {432, 1796}

\bibitem[\protect\citeauthoryear{{Allain}, {Leach}  \& {Sedlmayr}}{{Allain}
  et~al.}{1996}]{Allain96}
{Allain} T.,  {Leach} S.,   {Sedlmayr} E.,  1996, \aap, \href
  {http://adsabs.harvard.edu/abs/1996A%26A...305..602A} {305, 602}

\bibitem[\protect\citeauthoryear{{Allamandola}, {Tielens}  \&
  {Barker}}{{Allamandola} et~al.}{1989}]{Allamandola89}
{Allamandola} L.~J.,  {Tielens} A.~G.~G.~M.,   {Barker} J.~R.,  1989, \mn@doi
  [\apjs] {10.1086/191396}, \href
  {http://adsabs.harvard.edu/abs/1989ApJS...71..733A} {71, 733}

\bibitem[\protect\citeauthoryear{{Allamandola}, {Hudgins}  \&
  {Sandford}}{{Allamandola} et~al.}{1999}]{Allamandola99}
{Allamandola} L.~J.,  {Hudgins} D.~M.,   {Sandford} S.~A.,  1999, \mn@doi
  [\apjl] {10.1086/311843}, \href
  {http://adsabs.harvard.edu/abs/1999ApJ...511L.115A} {511, L115}

\bibitem[\protect\citeauthoryear{{Amblard}, {Riguccini}, {Temi}, {Im},
  {Fanelli}  \& {Serra}}{{Amblard} et~al.}{2014}]{Amblard14}
{Amblard} A.,  {Riguccini} L.,  {Temi} P.,  {Im} S.,  {Fanelli} M.,   {Serra}
  P.,  2014, \mn@doi [\apj] {10.1088/0004-637X/783/2/135}, \href
  {http://adsabs.harvard.edu/abs/2014ApJ...783..135A} {783, 135}

\bibitem[\protect\citeauthoryear{{Astropy Collaboration} et~al.,}{{Astropy
  Collaboration} et~al.}{2013}]{Astropy}
{Astropy Collaboration} et~al., 2013, \mn@doi [\aap]
  {10.1051/0004-6361/201322068}, \href
  {http://adsabs.harvard.edu/abs/2013A%26A...558A..33A} {558, A33}

\bibitem[\protect\citeauthoryear{{Athey}, {Bregman}, {Bregman}, {Temi}  \&
  {Sauvage}}{{Athey} et~al.}{2002}]{Athey02}
{Athey} A.,  {Bregman} J.,  {Bregman} J.,  {Temi} P.,   {Sauvage} M.,  2002,
  \mn@doi [\apj] {10.1086/339844}, \href
  {http://adsabs.harvard.edu/abs/2002ApJ...571..272A} {571, 272}

\bibitem[\protect\citeauthoryear{{Bai}, {Zou}, {Liu}  \& {Wang}}{{Bai}
  et~al.}{2015}]{Bai15}
{Bai} Y.,  {Zou} H.,  {Liu} J.,   {Wang} S.,  2015, \mn@doi [\apjs]
  {10.1088/0067-0049/220/1/6}, \href
  {http://adsabs.harvard.edu/abs/2015ApJS..220....6B} {220, 6}

\bibitem[\protect\citeauthoryear{{Bakes} \& {Tielens}}{{Bakes} \&
  {Tielens}}{1994}]{Bakes94}
{Bakes} E.~L.~O.,  {Tielens} A.~G.~G.~M.,  1994, \mn@doi [\apj]
  {10.1086/174188}, \href {http://adsabs.harvard.edu/abs/1994ApJ...427..822B}
  {427, 822}

\bibitem[\protect\citeauthoryear{{Bedijn}}{{Bedijn}}{1987}]{Bedijn87}
{Bedijn} P.~J.,  1987, \aap, \href
  {http://adsabs.harvard.edu/abs/1987A%26A...186..136B} {186, 136}

\bibitem[\protect\citeauthoryear{{Bell} \& {de Jong}}{{Bell} \& {de
  Jong}}{2001}]{Bell01}
{Bell} E.~F.,  {de Jong} R.~S.,  2001, \mn@doi [\apj] {10.1086/319728}, \href
  {http://adsabs.harvard.edu/abs/2001ApJ...550..212B} {550, 212}

\bibitem[\protect\citeauthoryear{{Bell}, {McIntosh}, {Katz}  \&
  {Weinberg}}{{Bell} et~al.}{2003}]{Bell03}
{Bell} E.~F.,  {McIntosh} D.~H.,  {Katz} N.,   {Weinberg} M.~D.,  2003, \mn@doi
  [\apjs] {10.1086/378847}, \href
  {http://adsabs.harvard.edu/abs/2003ApJS..149..289B} {149, 289}

\bibitem[\protect\citeauthoryear{{Bertin} \& {Arnouts}}{{Bertin} \&
  {Arnouts}}{1996}]{Bertin96}
{Bertin} E.,  {Arnouts} S.,  1996, \mn@doi [\aaps] {10.1051/aas:1996164}, \href
  {http://adsabs.harvard.edu/abs/1996A%26AS..117..393B} {117, 393}

\bibitem[\protect\citeauthoryear{{Blanton} \& {Roweis}}{{Blanton} \&
  {Roweis}}{2007}]{Blanton07}
{Blanton} M.~R.,  {Roweis} S.,  2007, \mn@doi [\aj] {10.1086/510127}, \href
  {http://adsabs.harvard.edu/abs/2007AJ....133..734B} {133, 734}

\bibitem[\protect\citeauthoryear{{Brandl} et~al.,}{{Brandl}
  et~al.}{2004}]{Brandl04}
{Brandl} B.~R.,  et~al., 2004, \mn@doi [\apjs] {10.1086/422101}, \href
  {http://adsabs.harvard.edu/abs/2004ApJS..154..188B} {154, 188}

\bibitem[\protect\citeauthoryear{{Bregman} \& {Temi}}{{Bregman} \&
  {Temi}}{2005}]{Bregman05}
{Bregman} J.,  {Temi} P.,  2005, \mn@doi [\apj] {10.1086/427738}, \href
  {http://adsabs.harvard.edu/abs/2005ApJ...621..831B} {621, 831}

\bibitem[\protect\citeauthoryear{{Bregman}, {Snider}, {Grego}  \&
  {Cox}}{{Bregman} et~al.}{1998}]{Bregman98}
{Bregman} J.~N.,  {Snider} B.~A.,  {Grego} L.,   {Cox} C.~V.,  1998, \apj,
  \href {http://adsabs.harvard.edu/abs/1998ApJ...499..670B} {499, 670}

\bibitem[\protect\citeauthoryear{{Bregman}, {Temi}  \& {Bregman}}{{Bregman}
  et~al.}{2006}]{Bregman06}
{Bregman} J.~N.,  {Temi} P.,   {Bregman} J.~D.,  2006, \mn@doi [\apj]
  {10.1086/505190}, \href {http://adsabs.harvard.edu/abs/2006ApJ...647..265B}
  {647, 265}

\bibitem[\protect\citeauthoryear{{Bressan}, {Granato}  \& {Silva}}{{Bressan}
  et~al.}{1998}]{Bressan98}
{Bressan} A.,  {Granato} G.~L.,   {Silva} L.,  1998, \aap, \href
  {http://adsabs.harvard.edu/abs/1998A%26A...332..135B} {332, 135}

\bibitem[\protect\citeauthoryear{{Bressan}, {Marigo}, {Girardi}, {Salasnich},
  {Dal Cero}, {Rubele}  \& {Nanni}}{{Bressan} et~al.}{2012}]{Bressan12}
{Bressan} A.,  {Marigo} P.,  {Girardi} L.,  {Salasnich} B.,  {Dal Cero} C.,
  {Rubele} S.,   {Nanni} A.,  2012, \mn@doi [\mnras]
  {10.1111/j.1365-2966.2012.21948.x}, \href
  {http://adsabs.harvard.edu/abs/2012MNRAS.427..127B} {427, 127}

\bibitem[\protect\citeauthoryear{{Bruzual} \& {Charlot}}{{Bruzual} \&
  {Charlot}}{2003}]{Bruzual03}
{Bruzual} G.,  {Charlot} S.,  2003, \mn@doi [\mnras]
  {10.1046/j.1365-8711.2003.06897.x}, \href
  {http://adsabs.harvard.edu/abs/2003MNRAS.344.1000B} {344, 1000}

\bibitem[\protect\citeauthoryear{{Caldwell}}{{Caldwell}}{1984}]{Caldwell84}
{Caldwell} N.,  1984, \mn@doi [\pasp] {10.1086/131334}, \href
  {http://adsabs.harvard.edu/abs/1984PASP...96..287C} {96, 287}

\bibitem[\protect\citeauthoryear{{Canizares}, {Fabbiano}  \&
  {Trinchieri}}{{Canizares} et~al.}{1987}]{Canizares87}
{Canizares} C.~R.,  {Fabbiano} G.,   {Trinchieri} G.,  1987, \mn@doi [\apj]
  {10.1086/164896}, \href {http://adsabs.harvard.edu/abs/1987ApJ...312..503C}
  {312, 503}

\bibitem[\protect\citeauthoryear{{Cappellari}}{{Cappellari}}{2008}]{Cappellari08}
{Cappellari} M.,  2008, \mn@doi [\mnras] {10.1111/j.1365-2966.2008.13754.x},
  \href {http://adsabs.harvard.edu/abs/2008MNRAS.390...71C} {390, 71}

\bibitem[\protect\citeauthoryear{{Cappellari} et~al.,}{{Cappellari}
  et~al.}{2011}]{Cappellari11}
{Cappellari} M.,  et~al., 2011, \mn@doi [\mnras]
  {10.1111/j.1365-2966.2010.18174.x}, \href
  {http://adsabs.harvard.edu/abs/2011MNRAS.413..813C} {413, 813}

\bibitem[\protect\citeauthoryear{{Cappellari} et~al.,}{{Cappellari}
  et~al.}{2012}]{Cappellari12}
{Cappellari} M.,  et~al., 2012, \mn@doi [\nat] {10.1038/nature10972}, \href
  {http://adsabs.harvard.edu/abs/2012Natur.484..485C} {484, 485}

\bibitem[\protect\citeauthoryear{{Cappellari} et~al.,}{{Cappellari}
  et~al.}{2013}]{Cappellari13a}
{Cappellari} M.,  et~al., 2013, \mn@doi [\mnras] {10.1093/mnras/stt562}, \href
  {http://adsabs.harvard.edu/abs/2013MNRAS.432.1709C} {432, 1709}

\bibitem[\protect\citeauthoryear{{Cassar{\`a}}, {Piovan}, {Weiss}, {Salaris}
  \& {Chiosi}}{{Cassar{\`a}} et~al.}{2013}]{Cassara13}
{Cassar{\`a}} L.~P.,  {Piovan} L.,  {Weiss} A.,  {Salaris} M.,   {Chiosi} C.,
  2013, \mn@doi [\mnras] {10.1093/mnras/stt1778}, \href
  {http://adsabs.harvard.edu/abs/2013MNRAS.436.2824C} {436, 2824}

\bibitem[\protect\citeauthoryear{{Chabrier}}{{Chabrier}}{2003}]{Chabrier03}
{Chabrier} G.,  2003, \mn@doi [\pasp] {10.1086/376392}, \href
  {http://adsabs.harvard.edu/abs/2003PASP..115..763C} {115, 763}

\bibitem[\protect\citeauthoryear{{Cluver} et~al.,}{{Cluver}
  et~al.}{2014}]{Cluver14}
{Cluver} M.~E.,  et~al., 2014, \mn@doi [\apj] {10.1088/0004-637X/782/2/90},
  \href {http://adsabs.harvard.edu/abs/2014ApJ...782...90C} {782, 90}

\bibitem[\protect\citeauthoryear{{Conroy}}{{Conroy}}{2013}]{Conroy13}
{Conroy} C.,  2013, \mn@doi [\araa] {10.1146/annurev-astro-082812-141017},
  \href {http://adsabs.harvard.edu/abs/2013ARA%26A..51..393C} {51, 393}

\bibitem[\protect\citeauthoryear{{Conroy} \& {Gunn}}{{Conroy} \&
  {Gunn}}{2010}]{Conroy10}
{Conroy} C.,  {Gunn} J.~E.,  2010, \mn@doi [\apj]
  {10.1088/0004-637X/712/2/833}, \href
  {http://adsabs.harvard.edu/abs/2010ApJ...712..833C} {712, 833}

\bibitem[\protect\citeauthoryear{{Conroy}, {Gunn}  \& {White}}{{Conroy}
  et~al.}{2009}]{Conroy09}
{Conroy} C.,  {Gunn} J.~E.,   {White} M.,  2009, \mn@doi [\apj]
  {10.1088/0004-637X/699/1/486}, \href
  {http://adsabs.harvard.edu/abs/2009ApJ...699..486C} {699, 486}

\bibitem[\protect\citeauthoryear{{Davis} et~al.,}{{Davis}
  et~al.}{2014}]{Davis14}
{Davis} T.~A.,  et~al., 2014, \mn@doi [\mnras] {10.1093/mnras/stu570}, \href
  {http://adsabs.harvard.edu/abs/2014MNRAS.444.3427D} {444, 3427}

\bibitem[\protect\citeauthoryear{{DeFrees}, {Miller}, {Talbi}, {Pauzat}  \&
  {Ellinger}}{{DeFrees} et~al.}{1993}]{DeFrees93}
{DeFrees} D.~J.,  {Miller} M.~D.,  {Talbi} D.,  {Pauzat} F.,   {Ellinger} Y.,
  1993, \mn@doi [\apj] {10.1086/172610}, \href
  {http://adsabs.harvard.edu/abs/1993ApJ...408..530D} {408, 530}

\bibitem[\protect\citeauthoryear{{Diamond-Stanic} \& {Rieke}}{{Diamond-Stanic}
  \& {Rieke}}{2010}]{Diamond10}
{Diamond-Stanic} A.~M.,  {Rieke} G.~H.,  2010, \mn@doi [\apj]
  {10.1088/0004-637X/724/1/140}, \href
  {http://adsabs.harvard.edu/abs/2010ApJ...724..140D} {724, 140}

\bibitem[\protect\citeauthoryear{{Draine} \& {Li}}{{Draine} \&
  {Li}}{2001}]{Draine01}
{Draine} B.~T.,  {Li} A.,  2001, \mn@doi [\apj] {10.1086/320227}, \href
  {http://adsabs.harvard.edu/abs/2001ApJ...551..807D} {551, 807}

\bibitem[\protect\citeauthoryear{{Draine} \& {Li}}{{Draine} \&
  {Li}}{2007}]{Draine07}
{Draine} B.~T.,  {Li} A.,  2007, \mn@doi [\apj] {10.1086/511055}, \href
  {http://adsabs.harvard.edu/abs/2007ApJ...657..810D} {657, 810}

\bibitem[\protect\citeauthoryear{{Draine} \& {Salpeter}}{{Draine} \&
  {Salpeter}}{1979}]{Draine79}
{Draine} B.~T.,  {Salpeter} E.~E.,  1979, \mn@doi [\apj] {10.1086/157206},
  \href {http://adsabs.harvard.edu/abs/1979ApJ...231..438D} {231, 438}

\bibitem[\protect\citeauthoryear{{Duley} \& {Williams}}{{Duley} \&
  {Williams}}{1981}]{Duley81}
{Duley} W.~W.,  {Williams} D.~A.,  1981, \mnras, \href
  {http://adsabs.harvard.edu/abs/1981MNRAS.196..269D} {196, 269}

\bibitem[\protect\citeauthoryear{{Dwek} \& {Arendt}}{{Dwek} \&
  {Arendt}}{1992}]{Dwek92}
{Dwek} E.,  {Arendt} R.~G.,  1992, \mn@doi [\araa]
  {10.1146/annurev.aa.30.090192.000303}, \href
  {http://adsabs.harvard.edu/abs/1992ARA%26A..30...11D} {30, 11}

\bibitem[\protect\citeauthoryear{{Ebneter}, {Davis}  \& {Djorgovski}}{{Ebneter}
  et~al.}{1988}]{Ebneter88}
{Ebneter} K.,  {Davis} M.,   {Djorgovski} S.,  1988, \mn@doi [\aj]
  {10.1086/114644}, \href {http://adsabs.harvard.edu/abs/1988AJ.....95..422E}
  {95, 422}

\bibitem[\protect\citeauthoryear{{Ferrari}, {Pastoriza}, {Macchetto},
  {Bonatto}, {Panagia}  \& {Sparks}}{{Ferrari} et~al.}{2002}]{Ferrari02}
{Ferrari} F.,  {Pastoriza} M.~G.,  {Macchetto} F.~D.,  {Bonatto} C.,  {Panagia}
  N.,   {Sparks} W.~B.,  2002, \mn@doi [\aap] {10.1051/0004-6361:20020582},
  \href {http://adsabs.harvard.edu/abs/2002A%26A...389..355F} {389, 355}

\bibitem[\protect\citeauthoryear{{Ford} \& {Bregman}}{{Ford} \&
  {Bregman}}{2013}]{Ford13}
{Ford} H.~A.,  {Bregman} J.~N.,  2013, \mn@doi [\apj]
  {10.1088/0004-637X/770/2/137}, \href
  {http://adsabs.harvard.edu/abs/2013ApJ...770..137F} {770, 137}

\bibitem[\protect\citeauthoryear{{Forman}, {Jones}  \& {Tucker}}{{Forman}
  et~al.}{1985}]{Forman85}
{Forman} W.,  {Jones} C.,   {Tucker} W.,  1985, \mn@doi [\apj]
  {10.1086/163218}, \href {http://adsabs.harvard.edu/abs/1985ApJ...293..102F}
  {293, 102}

\bibitem[\protect\citeauthoryear{{Galliano}, {Madden}, {Tielens}, {Peeters}  \&
  {Jones}}{{Galliano} et~al.}{2008}]{Galliano08}
{Galliano} F.,  {Madden} S.~C.,  {Tielens} A.~G.~G.~M.,  {Peeters} E.,
  {Jones} A.~P.,  2008, \mn@doi [\apj] {10.1086/587051}, \href
  {http://adsabs.harvard.edu/abs/2008ApJ...679..310G} {679, 310}

\bibitem[\protect\citeauthoryear{{Gil de Paz} et~al.,}{{Gil de Paz}
  et~al.}{2007}]{GildePaz07}
{Gil de Paz} A.,  et~al., 2007, \mn@doi [\apjs] {10.1086/516636}, \href
  {http://adsabs.harvard.edu/abs/2007ApJS..173..185G} {173, 185}

\bibitem[\protect\citeauthoryear{{Girardi} et~al.,}{{Girardi}
  et~al.}{2010}]{Girardi10}
{Girardi} L.,  et~al., 2010, \mn@doi [\apj] {10.1088/0004-637X/724/2/1030},
  \href {http://adsabs.harvard.edu/abs/2010ApJ...724.1030G} {724, 1030}

\bibitem[\protect\citeauthoryear{{Goldreich} \& {Scoville}}{{Goldreich} \&
  {Scoville}}{1976}]{Goldreich76}
{Goldreich} P.,  {Scoville} N.,  1976, \mn@doi [\apj] {10.1086/154257}, \href
  {http://adsabs.harvard.edu/abs/1976ApJ...205..144G} {205, 144}

\bibitem[\protect\citeauthoryear{{Goudfrooij} \& {de Jong}}{{Goudfrooij} \& {de
  Jong}}{1995}]{Goudfrooij95}
{Goudfrooij} P.,  {de Jong} T.,  1995, \aap, \href
  {http://adsabs.harvard.edu/abs/1995A%26A...298..784G} {298, 784}

\bibitem[\protect\citeauthoryear{{Goudfrooij}, {de Jong}, {Hansen}  \&
  {Norgaard-Nielsen}}{{Goudfrooij} et~al.}{1994}]{Goudfrooij94b}
{Goudfrooij} P.,  {de Jong} T.,  {Hansen} L.,   {Norgaard-Nielsen} H.~U.,
  1994, \mnras, \href {http://adsabs.harvard.edu/abs/1994MNRAS.271..833G} {271,
  833}

\bibitem[\protect\citeauthoryear{{Groves} et~al.,}{{Groves}
  et~al.}{2012}]{Groves12}
{Groves} B.,  et~al., 2012, \mn@doi [\mnras]
  {10.1111/j.1365-2966.2012.21696.x}, \href
  {http://adsabs.harvard.edu/abs/2012MNRAS.426..892G} {426, 892}

\bibitem[\protect\citeauthoryear{{Helou}}{{Helou}}{1986}]{Helou86}
{Helou} G.,  1986, \mn@doi [\apjl] {10.1086/184793}, \href
  {http://adsabs.harvard.edu/abs/1986ApJ...311L..33H} {311, L33}

\bibitem[\protect\citeauthoryear{{Helou}, {Lu}, {Werner}, {Malhotra}  \&
  {Silbermann}}{{Helou} et~al.}{2000}]{Helou00}
{Helou} G.,  {Lu} N.~Y.,  {Werner} M.~W.,  {Malhotra} S.,   {Silbermann} N.,
  2000, \mn@doi [\apjl] {10.1086/312549}, \href
  {http://adsabs.harvard.edu/abs/2000ApJ...532L..21H} {532, L21}

\bibitem[\protect\citeauthoryear{{Ho}, {Filippenko}  \& {Sargent}}{{Ho}
  et~al.}{1997}]{Ho97}
{Ho} L.~C.,  {Filippenko} A.~V.,   {Sargent} W.~L.~W.,  1997, \mn@doi [\apjs]
  {10.1086/313041}, \href {http://adsabs.harvard.edu/abs/1997ApJS..112..315H}
  {112, 315}

\bibitem[\protect\citeauthoryear{{Hony}, {Van Kerckhoven}, {Peeters},
  {Tielens}, {Hudgins}  \& {Allamandola}}{{Hony} et~al.}{2001}]{Hony01}
{Hony} S.,  {Van Kerckhoven} C.,  {Peeters} E.,  {Tielens} A.~G.~G.~M.,
  {Hudgins} D.~M.,   {Allamandola} L.~J.,  2001, \mn@doi [\aap]
  {10.1051/0004-6361:20010242}, \href
  {http://adsabs.harvard.edu/abs/2001A%26A...370.1030H} {370, 1030}

\bibitem[\protect\citeauthoryear{Hudgins \& Allamandola}{Hudgins \&
  Allamandola}{1995}]{Hudgins95}
Hudgins D.~M.,  Allamandola L.~J.,  1995, \mn@doi [The Journal of Physical
  Chemistry] {10.1021/j100010a011}, 99, 3033

\bibitem[\protect\citeauthoryear{Hunter}{Hunter}{2007}]{Matplotlib}
Hunter J.~D.,  2007, Computing In Science \& Engineering, 9, 90

\bibitem[\protect\citeauthoryear{{Indebetouw} et~al.,}{{Indebetouw}
  et~al.}{2005}]{Indebetouw05}
{Indebetouw} R.,  et~al., 2005, \mn@doi [\apj] {10.1086/426679}, \href
  {http://adsabs.harvard.edu/abs/2005ApJ...619..931I} {619, 931}

\bibitem[\protect\citeauthoryear{{Jarrett} et~al.,}{{Jarrett}
  et~al.}{2011}]{Jarrett11}
{Jarrett} T.~H.,  et~al., 2011, \mn@doi [\apj] {10.1088/0004-637X/735/2/112},
  \href {http://adsabs.harvard.edu/abs/2011ApJ...735..112J} {735, 112}

\bibitem[\protect\citeauthoryear{{Jarrett} et~al.,}{{Jarrett}
  et~al.}{2013}]{Jarrett13}
{Jarrett} T.~H.,  et~al., 2013, \mn@doi [\aj] {10.1088/0004-6256/145/1/6},
  \href {http://adsabs.harvard.edu/abs/2013AJ....145....6J} {145, 6}

\bibitem[\protect\citeauthoryear{{Joblin}, {Tielens}, {Geballe}  \&
  {Wooden}}{{Joblin} et~al.}{1996}]{Joblin96}
{Joblin} C.,  {Tielens} A.~G.~G.~M.,  {Geballe} T.~R.,   {Wooden} D.~H.,  1996,
  \mn@doi [\apjl] {10.1086/309986}, \href
  {http://adsabs.harvard.edu/abs/1996ApJ...460L.119J} {460, L119}

\bibitem[\protect\citeauthoryear{{Jochims}, {Ruhl}, {Baumgartel}, {Tobita}  \&
  {Leach}}{{Jochims} et~al.}{1994}]{Jochims94}
{Jochims} H.~W.,  {Ruhl} E.,  {Baumgartel} H.,  {Tobita} S.,   {Leach} S.,
  1994, \mn@doi [\apj] {10.1086/173560}, \href
  {http://adsabs.harvard.edu/abs/1994ApJ...420..307J} {420, 307}

\bibitem[\protect\citeauthoryear{{Jones}, {Tielens}, {Hollenbach}  \&
  {McKee}}{{Jones} et~al.}{1994}]{Jones94}
{Jones} A.~P.,  {Tielens} A.~G.~G.~M.,  {Hollenbach} D.~J.,   {McKee} C.~F.,
  1994, \mn@doi [\apj] {10.1086/174689}, \href
  {http://adsabs.harvard.edu/abs/1994ApJ...433..797J} {433, 797}

\bibitem[\protect\citeauthoryear{Jones, Oliphant, Peterson  et~al.}{Jones
  et~al.}{01  }]{Scipy}
Jones E.,  Oliphant T.,  Peterson P.,   et~al., 2001--, {SciPy}: Open source
  scientific tools for {Python}, \url {http://www.scipy.org/}

\bibitem[\protect\citeauthoryear{{Jura}, {Kim}, {Knapp}  \&
  {Guhathakurta}}{{Jura} et~al.}{1987}]{Jura87}
{Jura} M.,  {Kim} D.~W.,  {Knapp} G.~R.,   {Guhathakurta} P.,  1987, \mn@doi
  [\apjl] {10.1086/184810}, \href
  {http://adsabs.harvard.edu/abs/1987ApJ...312L..11J} {312, L11}

\bibitem[\protect\citeauthoryear{{Kaneda}, {Onaka}  \& {Sakon}}{{Kaneda}
  et~al.}{2005}]{Kaneda05}
{Kaneda} H.,  {Onaka} T.,   {Sakon} I.,  2005, \mn@doi [\apjl]
  {10.1086/497913}, \href {http://adsabs.harvard.edu/abs/2005ApJ...632L..83K}
  {632, L83}

\bibitem[\protect\citeauthoryear{{Kaneda}, {Onaka}, {Sakon}, {Kitayama},
  {Okada}  \& {Suzuki}}{{Kaneda} et~al.}{2008}]{Kaneda08}
{Kaneda} H.,  {Onaka} T.,  {Sakon} I.,  {Kitayama} T.,  {Okada} Y.,   {Suzuki}
  T.,  2008, \mn@doi [\apj] {10.1086/590243}, \href
  {http://adsabs.harvard.edu/abs/2008ApJ...684..270K} {684, 270}

\bibitem[\protect\citeauthoryear{{Kaviraj} et~al.,}{{Kaviraj}
  et~al.}{2007}]{Kaviraj07}
{Kaviraj} S.,  et~al., 2007, \mn@doi [\apjs] {10.1086/516633}, \href
  {http://adsabs.harvard.edu/abs/2007ApJS..173..619K} {173, 619}

\bibitem[\protect\citeauthoryear{{Kelson} \& {Holden}}{{Kelson} \&
  {Holden}}{2010}]{Kelson10}
{Kelson} D.~D.,  {Holden} B.~P.,  2010, \mn@doi [\apjl]
  {10.1088/2041-8205/713/1/L28}, \href
  {http://adsabs.harvard.edu/abs/2010ApJ...713L..28K} {713, L28}

\bibitem[\protect\citeauthoryear{{Kennicutt}}{{Kennicutt}}{1998a}]{Kennicutt98b}
{Kennicutt} Jr. R.~C.,  1998a, \mn@doi [\araa]
  {10.1146/annurev.astro.36.1.189}, \href
  {http://adsabs.harvard.edu/abs/1998ARA%26A..36..189K} {36, 189}

\bibitem[\protect\citeauthoryear{{Kennicutt}}{{Kennicutt}}{1998b}]{Kennicutt98a}
{Kennicutt} Jr. R.~C.,  1998b, \mn@doi [\apj] {10.1086/305588}, \href
  {http://adsabs.harvard.edu/abs/1998ApJ...498..541K} {498, 541}

\bibitem[\protect\citeauthoryear{{Kennicutt} Jr. et~al.,}{{Kennicutt}
  et~al.}{2003}]{Kennicutt03}
{Kennicutt} Jr. R.~C.,  et~al., 2003, \mn@doi [\pasp] {10.1086/376941}, \href
  {http://adsabs.harvard.edu/abs/2003PASP..115..928K} {115, 928}

\bibitem[\protect\citeauthoryear{{Knapp}, {Turner}  \& {Cunniffe}}{{Knapp}
  et~al.}{1985}]{Knapp85}
{Knapp} G.~R.,  {Turner} E.~L.,   {Cunniffe} P.~E.,  1985, \mn@doi [\aj]
  {10.1086/113751}, \href {http://adsabs.harvard.edu/abs/1985AJ.....90..454K}
  {90, 454}

\bibitem[\protect\citeauthoryear{{Knapp}, {Guhathakurta}, {Kim}  \&
  {Jura}}{{Knapp} et~al.}{1989}]{Knapp89}
{Knapp} G.~R.,  {Guhathakurta} P.,  {Kim} D.-W.,   {Jura} M.~A.,  1989, \mn@doi
  [\apjs] {10.1086/191342}, \href
  {http://adsabs.harvard.edu/abs/1989ApJS...70..329K} {70, 329}

\bibitem[\protect\citeauthoryear{{Krajnovi{\'c}} et~al.,}{{Krajnovi{\'c}}
  et~al.}{2011}]{Krajnovic11}
{Krajnovi{\'c}} D.,  et~al., 2011, \mn@doi [\mnras]
  {10.1111/j.1365-2966.2011.18560.x}, \href
  {http://adsabs.harvard.edu/abs/2011MNRAS.414.2923K} {414, 2923}

\bibitem[\protect\citeauthoryear{Langhoff}{Langhoff}{1996}]{Langhoff96}
Langhoff S.~R.,  1996, \mn@doi [The Journal of Physical Chemistry]
  {10.1021/jp952074g}, 100, 2819

\bibitem[\protect\citeauthoryear{{Lonsdale Persson} \& {Helou}}{{Lonsdale
  Persson} \& {Helou}}{1987}]{Lonsdale87}
{Lonsdale Persson} C.~J.,  {Helou} G.,  1987, \mn@doi [\apj] {10.1086/165082},
  \href {http://adsabs.harvard.edu/abs/1987ApJ...314..513L} {314, 513}

\bibitem[\protect\citeauthoryear{{MacArthur}, {Courteau}, {Bell}  \&
  {Holtzman}}{{MacArthur} et~al.}{2004}]{MacArthur04}
{MacArthur} L.~A.,  {Courteau} S.,  {Bell} E.,   {Holtzman} J.~A.,  2004,
  \mn@doi [\apjs] {10.1086/383525}, \href
  {http://adsabs.harvard.edu/abs/2004ApJS..152..175M} {152, 175}

\bibitem[\protect\citeauthoryear{{Madden}, {Vigroux}  \& {Sauvage}}{{Madden}
  et~al.}{1999}]{Madden99}
{Madden} S.~C.,  {Vigroux} L.,   {Sauvage} M.,  1999, in {Cox} P.,  {Kessler}
  M.,  eds,  ESA Special Publication Vol. 427, The Universe as Seen by ISO.
  p.~933

\bibitem[\protect\citeauthoryear{{Maraston}}{{Maraston}}{2005}]{Maraston05}
{Maraston} C.,  2005, \mn@doi [\mnras] {10.1111/j.1365-2966.2005.09270.x},
  \href {http://adsabs.harvard.edu/abs/2005MNRAS.362..799M} {362, 799}

\bibitem[\protect\citeauthoryear{{Maraston}, {Daddi}, {Renzini}, {Cimatti},
  {Dickinson}, {Papovich}, {Pasquali}  \& {Pirzkal}}{{Maraston}
  et~al.}{2006}]{Maraston06}
{Maraston} C.,  {Daddi} E.,  {Renzini} A.,  {Cimatti} A.,  {Dickinson} M.,
  {Papovich} C.,  {Pasquali} A.,   {Pirzkal} N.,  2006, \mn@doi [\apj]
  {10.1086/508143}, \href {http://adsabs.harvard.edu/abs/2006ApJ...652...85M}
  {652, 85}

\bibitem[\protect\citeauthoryear{{Marigo}, {Girardi}, {Bressan}, {Groenewegen},
  {Silva}  \& {Granato}}{{Marigo} et~al.}{2008}]{Marigo08}
{Marigo} P.,  {Girardi} L.,  {Bressan} A.,  {Groenewegen} M.~A.~T.,  {Silva}
  L.,   {Granato} G.~L.,  2008, \mn@doi [\aap] {10.1051/0004-6361:20078467},
  \href {http://adsabs.harvard.edu/abs/2008A%26A...482..883M} {482, 883}

\bibitem[\protect\citeauthoryear{{Marigo}, {Bressan}, {Nanni}, {Girardi}  \&
  {Pumo}}{{Marigo} et~al.}{2013}]{Marigo13}
{Marigo} P.,  {Bressan} A.,  {Nanni} A.,  {Girardi} L.,   {Pumo} M.~L.,  2013,
  \mn@doi [\mnras] {10.1093/mnras/stt1034}, \href
  {http://adsabs.harvard.edu/abs/2013MNRAS.434..488M} {434, 488}

\bibitem[\protect\citeauthoryear{{Martin} et~al.,}{{Martin}
  et~al.}{2005}]{Martin05}
{Martin} D.~C.,  et~al., 2005, \mn@doi [\apjl] {10.1086/426387}, \href
  {http://adsabs.harvard.edu/abs/2005ApJ...619L...1M} {619, L1}

\bibitem[\protect\citeauthoryear{{Martini}, {Dicken}  \&
  {Storchi-Bergmann}}{{Martini} et~al.}{2013}]{Martini13}
{Martini} P.,  {Dicken} D.,   {Storchi-Bergmann} T.,  2013, \mn@doi [\apj]
  {10.1088/0004-637X/766/2/121}, \href
  {http://adsabs.harvard.edu/abs/2013ApJ...766..121M} {766, 121}

\bibitem[\protect\citeauthoryear{{McDermid} et~al.,}{{McDermid}
  et~al.}{2015}]{McDermid15}
{McDermid} R.~M.,  et~al., 2015, \mn@doi [\mnras] {10.1093/mnras/stv105}, \href
  {http://adsabs.harvard.edu/abs/2015MNRAS.448.3484M} {448, 3484}

\bibitem[\protect\citeauthoryear{{Meidt} et~al.,}{{Meidt}
  et~al.}{2012}]{Meidt12}
{Meidt} S.~E.,  et~al., 2012, \mn@doi [\apj] {10.1088/0004-637X/744/1/17},
  \href {http://adsabs.harvard.edu/abs/2012ApJ...744...17M} {744, 17}

\bibitem[\protect\citeauthoryear{{Meidt} et~al.,}{{Meidt}
  et~al.}{2014}]{Meidt14}
{Meidt} S.~E.,  et~al., 2014, \mn@doi [\apj] {10.1088/0004-637X/788/2/144},
  \href {http://adsabs.harvard.edu/abs/2014ApJ...788..144M} {788, 144}

\bibitem[\protect\citeauthoryear{{Melbourne} \& {Boyer}}{{Melbourne} \&
  {Boyer}}{2013}]{Melbourne13}
{Melbourne} J.,  {Boyer} M.~L.,  2013, \mn@doi [\apj]
  {10.1088/0004-637X/764/1/30}, \href
  {http://adsabs.harvard.edu/abs/2013ApJ...764...30M} {764, 30}

\bibitem[\protect\citeauthoryear{{Melbourne} et~al.,}{{Melbourne}
  et~al.}{2012}]{Melbourne12}
{Melbourne} J.,  et~al., 2012, \mn@doi [\apj] {10.1088/0004-637X/748/1/47},
  \href {http://adsabs.harvard.edu/abs/2012ApJ...748...47M} {748, 47}

\bibitem[\protect\citeauthoryear{{Micelotta}, {Jones}  \&
  {Tielens}}{{Micelotta} et~al.}{2010}]{Micelotta10}
{Micelotta} E.~R.,  {Jones} A.~P.,   {Tielens} A.~G.~G.~M.,  2010, \mn@doi
  [\aap] {10.1051/0004-6361/200911682}, \href
  {http://adsabs.harvard.edu/abs/2010A%26A...510A..36M} {510, A36}

\bibitem[\protect\citeauthoryear{{Morrissey} et~al.,}{{Morrissey}
  et~al.}{2005}]{Morrissey05}
{Morrissey} P.,  et~al., 2005, \mn@doi [\apjl] {10.1086/424734}, \href
  {http://adsabs.harvard.edu/abs/2005ApJ...619L...7M} {619, L7}

\bibitem[\protect\citeauthoryear{{Morrissey} et~al.,}{{Morrissey}
  et~al.}{2007}]{Morrissey07}
{Morrissey} P.,  et~al., 2007, \mn@doi [\apjs] {10.1086/520512}, \href
  {http://adsabs.harvard.edu/abs/2007ApJS..173..682M} {173, 682}

\bibitem[\protect\citeauthoryear{{Navarro}, {Frenk}  \& {White}}{{Navarro}
  et~al.}{1996}]{Navarro96}
{Navarro} J.~F.,  {Frenk} C.~S.,   {White} S.~D.~M.,  1996, \mn@doi [\apj]
  {10.1086/177173}, \href {http://adsabs.harvard.edu/abs/1996ApJ...462..563N}
  {462, 563}

\bibitem[\protect\citeauthoryear{{Norris}, {Meidt}, {Van de Ven}, {Schinnerer},
  {Groves}  \& {Querejeta}}{{Norris} et~al.}{2014}]{Norris14}
{Norris} M.~A.,  {Meidt} S.,  {Van de Ven} G.,  {Schinnerer} E.,  {Groves} B.,
   {Querejeta} M.,  2014, \mn@doi [\apj] {10.1088/0004-637X/797/1/55}, \href
  {http://adsabs.harvard.edu/abs/2014ApJ...797...55N} {797, 55}

\bibitem[\protect\citeauthoryear{{O'Dowd} et~al.,}{{O'Dowd}
  et~al.}{2009}]{Odowd09}
{O'Dowd} M.~J.,  et~al., 2009, \mn@doi [\apj] {10.1088/0004-637X/705/1/885},
  \href {http://adsabs.harvard.edu/abs/2009ApJ...705..885O} {705, 885}

\bibitem[\protect\citeauthoryear{{Ogle}, {Antonucci}, {Appleton}  \&
  {Whysong}}{{Ogle} et~al.}{2007}]{Ogle07}
{Ogle} P.,  {Antonucci} R.,  {Appleton} P.~N.,   {Whysong} D.,  2007, \mn@doi
  [\apj] {10.1086/521334}, \href
  {http://adsabs.harvard.edu/abs/2007ApJ...668..699O} {668, 699}

\bibitem[\protect\citeauthoryear{{Pahre}, {Ashby}, {Fazio}  \&
  {Willner}}{{Pahre} et~al.}{2004}]{Pahre04}
{Pahre} M.~A.,  {Ashby} M.~L.~N.,  {Fazio} G.~G.,   {Willner} S.~P.,  2004,
  \mn@doi [\apjs] {10.1086/423320}, \href
  {http://adsabs.harvard.edu/abs/2004ApJS..154..229P} {154, 229}

\bibitem[\protect\citeauthoryear{{Panuzzo}, {Rampazzo}, {Bressan}, {Vega},
  {Annibali}, {Buson}, {Clemens}  \& {Zeilinger}}{{Panuzzo}
  et~al.}{2011}]{Panuzzo11}
{Panuzzo} P.,  {Rampazzo} R.,  {Bressan} A.,  {Vega} O.,  {Annibali} F.,
  {Buson} L.~M.,  {Clemens} M.~S.,   {Zeilinger} W.~W.,  2011, \mn@doi [\aap]
  {10.1051/0004-6361/201015908}, \href
  {http://adsabs.harvard.edu/abs/2011A%26A...528A..10P} {528, A10}

\bibitem[\protect\citeauthoryear{{Peletier} et~al.,}{{Peletier}
  et~al.}{2012}]{Peletier12}
{Peletier} R.~F.,  et~al., 2012, \mn@doi [\mnras]
  {10.1111/j.1365-2966.2011.19855.x}, \href
  {http://adsabs.harvard.edu/abs/2012MNRAS.419.2031P} {419, 2031}

\bibitem[\protect\citeauthoryear{{Pellegrini}}{{Pellegrini}}{2010}]{Pellegrini10}
{Pellegrini} S.,  2010, \mn@doi [\apj] {10.1088/0004-637X/717/2/640}, \href
  {http://adsabs.harvard.edu/abs/2010ApJ...717..640P} {717, 640}

\bibitem[\protect\citeauthoryear{P\'erez \& Granger}{P\'erez \&
  Granger}{2007}]{IPython}
P\'erez F.,  Granger B.~E.,  2007, \mn@doi [Computing in Science and
  Engineering] {10.1109/MCSE.2007.53}, 9, 21

\bibitem[\protect\citeauthoryear{{Phillips}, {Jenkins}, {Dopita}, {Sadler}  \&
  {Binette}}{{Phillips} et~al.}{1986}]{Phillips86}
{Phillips} M.~M.,  {Jenkins} C.~R.,  {Dopita} M.~A.,  {Sadler} E.~M.,
  {Binette} L.,  1986, \mn@doi [\aj] {10.1086/114083}, \href
  {http://adsabs.harvard.edu/abs/1986AJ.....91.1062P} {91, 1062}

\bibitem[\protect\citeauthoryear{{Planck Collaboration} et~al.,}{{Planck
  Collaboration} et~al.}{2014}]{Planck14}
{Planck Collaboration} et~al., 2014, \mn@doi [\aap]
  {10.1051/0004-6361/201321529}, \href
  {http://adsabs.harvard.edu/abs/2014A%26A...571A...1P} {571, A1}

\bibitem[\protect\citeauthoryear{{Rampazzo}, {Panuzzo}, {Vega}, {Marino},
  {Bressan}  \& {Clemens}}{{Rampazzo} et~al.}{2013}]{Rampazzo13}
{Rampazzo} R.,  {Panuzzo} P.,  {Vega} O.,  {Marino} A.,  {Bressan} A.,
  {Clemens} M.~S.,  2013, \mn@doi [\mnras] {10.1093/mnras/stt475}, \href
  {http://adsabs.harvard.edu/abs/2013MNRAS.432..374R} {432, 374}

\bibitem[\protect\citeauthoryear{{Rosenfield} et~al.,}{{Rosenfield}
  et~al.}{2014}]{Rosenfield14}
{Rosenfield} P.,  et~al., 2014, \mn@doi [\apj] {10.1088/0004-637X/790/1/22},
  \href {http://adsabs.harvard.edu/abs/2014ApJ...790...22R} {790, 22}

\bibitem[\protect\citeauthoryear{{Rosenfield}, {Marigo}, {Girardi},
  {Dalcanton}, {Bressan}, {Williams}  \& {Dolphin}}{{Rosenfield}
  et~al.}{2016}]{Rosenfield16}
{Rosenfield} P.,  {Marigo} P.,  {Girardi} L.,  {Dalcanton} J.~J.,  {Bressan}
  A.,  {Williams} B.~F.,   {Dolphin} A.,  2016, \mn@doi [\apj]
  {10.3847/0004-637X/822/2/73}, \href
  {http://adsabs.harvard.edu/abs/2016ApJ...822...73R} {822, 73}

\bibitem[\protect\citeauthoryear{{Roussel} et~al.,}{{Roussel}
  et~al.}{2007}]{Roussel07}
{Roussel} H.,  et~al., 2007, \mn@doi [\apj] {10.1086/521667}, \href
  {http://adsabs.harvard.edu/abs/2007ApJ...669..959R} {669, 959}

\bibitem[\protect\citeauthoryear{{Sadler}}{{Sadler}}{1987}]{Sadler87}
{Sadler} E.~M.,  1987, in {de Zeeuw} P.~T.,  ed.,  IAU Symposium Vol. 127,
  Structure and Dynamics of Elliptical Galaxies. pp 125--132

\bibitem[\protect\citeauthoryear{{Sadler} \& {Gerhard}}{{Sadler} \&
  {Gerhard}}{1985}]{Sadler85}
{Sadler} E.~M.,  {Gerhard} O.~E.,  1985, \mnras, \href
  {http://adsabs.harvard.edu/abs/1985MNRAS.214..177S} {214, 177}

\bibitem[\protect\citeauthoryear{{Salim} et~al.,}{{Salim}
  et~al.}{2007}]{Salim07}
{Salim} S.,  et~al., 2007, \mn@doi [\apjs] {10.1086/519218}, \href
  {http://adsabs.harvard.edu/abs/2007ApJS..173..267S} {173, 267}

\bibitem[\protect\citeauthoryear{{Salpeter}}{{Salpeter}}{1955}]{Salpeter55}
{Salpeter} E.~E.,  1955, \mn@doi [\apj] {10.1086/145971}, \href
  {http://adsabs.harvard.edu/abs/1955ApJ...121..161S} {121, 161}

\bibitem[\protect\citeauthoryear{{Salpeter}}{{Salpeter}}{1974a}]{Salpeter74a}
{Salpeter} E.~E.,  1974a, \mn@doi [\apj] {10.1086/153195}, \href
  {http://adsabs.harvard.edu/abs/1974ApJ...193..579S} {193, 579}

\bibitem[\protect\citeauthoryear{{Salpeter}}{{Salpeter}}{1974b}]{Salpeter74b}
{Salpeter} E.~E.,  1974b, \mn@doi [\apj] {10.1086/153196}, \href
  {http://adsabs.harvard.edu/abs/1974ApJ...193..585S} {193, 585}

\bibitem[\protect\citeauthoryear{{Sarzi} et~al.,}{{Sarzi}
  et~al.}{2010}]{Sarzi10}
{Sarzi} M.,  et~al., 2010, \mn@doi [\mnras] {10.1111/j.1365-2966.2009.16039.x},
  \href {http://adsabs.harvard.edu/abs/2010MNRAS.402.2187S} {402, 2187}

\bibitem[\protect\citeauthoryear{{Schlegel}, {Finkbeiner}  \&
  {Davis}}{{Schlegel} et~al.}{1998}]{Schlegel98}
{Schlegel} D.~J.,  {Finkbeiner} D.~P.,   {Davis} M.,  1998, \mn@doi [\apj]
  {10.1086/305772}, \href {http://adsabs.harvard.edu/abs/1998ApJ...500..525S}
  {500, 525}

\bibitem[\protect\citeauthoryear{{Schutte}, {Tielens}  \&
  {Allamandola}}{{Schutte} et~al.}{1993}]{Schutte93}
{Schutte} W.~A.,  {Tielens} A.~G.~G.~M.,   {Allamandola} L.~J.,  1993, \mn@doi
  [\apj] {10.1086/173173}, \href
  {http://adsabs.harvard.edu/abs/1993ApJ...415..397S} {415, 397}

\bibitem[\protect\citeauthoryear{{Scott} et~al.,}{{Scott}
  et~al.}{2013}]{Scott13}
{Scott} N.,  et~al., 2013, \mn@doi [\mnras] {10.1093/mnras/sts422}, \href
  {http://adsabs.harvard.edu/abs/2013MNRAS.432.1894S} {432, 1894}

\bibitem[\protect\citeauthoryear{{Serra} et~al.,}{{Serra}
  et~al.}{2012}]{Serra12}
{Serra} P.,  et~al., 2012, \mn@doi [\mnras] {10.1111/j.1365-2966.2012.20219.x},
  \href {http://adsabs.harvard.edu/abs/2012MNRAS.422.1835S} {422, 1835}

\bibitem[\protect\citeauthoryear{{Shapiro} et~al.,}{{Shapiro}
  et~al.}{2010}]{Shapiro10}
{Shapiro} K.~L.,  et~al., 2010, \mn@doi [\mnras]
  {10.1111/j.1365-2966.2009.16111.x}, \href
  {http://adsabs.harvard.edu/abs/2010MNRAS.402.2140S} {402, 2140}

\bibitem[\protect\citeauthoryear{{Silva}, {Granato}, {Bressan}  \&
  {Danese}}{{Silva} et~al.}{1998}]{Silva98}
{Silva} L.,  {Granato} G.~L.,  {Bressan} A.,   {Danese} L.,  1998, \mn@doi
  [\apj] {10.1086/306476}, \href
  {http://adsabs.harvard.edu/abs/1998ApJ...509..103S} {509, 103}

\bibitem[\protect\citeauthoryear{{Smith}, {Struck}, {Hancock}, {Appleton},
  {Charmandaris}  \& {Reach}}{{Smith} et~al.}{2007}]{Smith07}
{Smith} B.~J.,  {Struck} C.,  {Hancock} M.,  {Appleton} P.~N.,  {Charmandaris}
  V.,   {Reach} W.~T.,  2007, \mn@doi [\aj] {10.1086/510350}, \href
  {http://adsabs.harvard.edu/abs/2007AJ....133..791S} {133, 791}

\bibitem[\protect\citeauthoryear{{Smith} et~al.,}{{Smith}
  et~al.}{2012}]{Smith12}
{Smith} M.~W.~L.,  et~al., 2012, \mn@doi [\apj] {10.1088/0004-637X/748/2/123},
  \href {http://adsabs.harvard.edu/abs/2012ApJ...748..123S} {748, 123}

\bibitem[\protect\citeauthoryear{{Sparks}, {Wall}, {Thorne}, {Jorden}, {van
  Breda}, {Rudd}  \& {Jorgensen}}{{Sparks} et~al.}{1985}]{Sparks85}
{Sparks} W.~B.,  {Wall} J.~V.,  {Thorne} D.~J.,  {Jorden} P.~R.,  {van Breda}
  I.~G.,  {Rudd} P.~J.,   {Jorgensen} H.~E.,  1985, \mnras, \href
  {http://adsabs.harvard.edu/abs/1985MNRAS.217...87S} {217, 87}

\bibitem[\protect\citeauthoryear{{Szczepanski} \& {Vala}}{{Szczepanski} \&
  {Vala}}{1993}]{Szczepanski93}
{Szczepanski} J.,  {Vala} M.,  1993, \mn@doi [\apj] {10.1086/173110}, \href
  {http://adsabs.harvard.edu/abs/1993ApJ...414..646S} {414, 646}

\bibitem[\protect\citeauthoryear{{Taylor} et~al.,}{{Taylor}
  et~al.}{2011}]{Taylor11}
{Taylor} E.~N.,  et~al., 2011, \mn@doi [\mnras]
  {10.1111/j.1365-2966.2011.19536.x}, \href
  {http://adsabs.harvard.edu/abs/2011MNRAS.418.1587T} {418, 1587}

\bibitem[\protect\citeauthoryear{{Temi}, {Brighenti}  \& {Mathews}}{{Temi}
  et~al.}{2009}]{Temi09}
{Temi} P.,  {Brighenti} F.,   {Mathews} W.~G.,  2009, \mn@doi [\apj]
  {10.1088/0004-637X/695/1/1}, \href
  {http://adsabs.harvard.edu/abs/2009ApJ...695....1T} {695, 1}

\bibitem[\protect\citeauthoryear{{Trams} et~al.,}{{Trams}
  et~al.}{1999}]{Trams99}
{Trams} N.~R.,  et~al., 1999, \aap, \href
  {http://adsabs.harvard.edu/abs/1999A%26A...346..843T} {346, 843}

\bibitem[\protect\citeauthoryear{{Vega} et~al.,}{{Vega} et~al.}{2010}]{Vega10}
{Vega} O.,  et~al., 2010, \mn@doi [\apj] {10.1088/0004-637X/721/2/1090}, \href
  {http://adsabs.harvard.edu/abs/2010ApJ...721.1090V} {721, 1090}

\bibitem[\protect\citeauthoryear{{Veron-Cetty} \& {Veron}}{{Veron-Cetty} \&
  {Veron}}{1988}]{Veron88}
{Veron-Cetty} M.-P.,  {Veron} P.,  1988, \aap, \href
  {http://adsabs.harvard.edu/abs/1988A%26A...204...28V} {204, 28}

\bibitem[\protect\citeauthoryear{{Villaume}, {Conroy}  \& {Johnson}}{{Villaume}
  et~al.}{2015}]{Villaume15}
{Villaume} A.,  {Conroy} C.,   {Johnson} B.~D.,  2015, \mn@doi [\apj]
  {10.1088/0004-637X/806/1/82}, \href
  {http://adsabs.harvard.edu/abs/2015ApJ...806...82V} {806, 82}

\bibitem[\protect\citeauthoryear{{Wardle} \& {Knapp}}{{Wardle} \&
  {Knapp}}{1986}]{Wardle86}
{Wardle} M.,  {Knapp} G.~R.,  1986, \mn@doi [\aj] {10.1086/113976}, \href
  {http://adsabs.harvard.edu/abs/1986AJ.....91...23W} {91, 23}

\bibitem[\protect\citeauthoryear{{Weingartner} \& {Draine}}{{Weingartner} \&
  {Draine}}{2001}]{Weingartner01}
{Weingartner} J.~C.,  {Draine} B.~T.,  2001, \mn@doi [\apjs] {10.1086/320852},
  \href {http://adsabs.harvard.edu/abs/2001ApJS..134..263W} {134, 263}

\bibitem[\protect\citeauthoryear{{Wright} et~al.,}{{Wright}
  et~al.}{2010}]{Wright10}
{Wright} E.~L.,  et~al., 2010, \mn@doi [\aj] {10.1088/0004-6256/140/6/1868},
  \href {http://adsabs.harvard.edu/abs/2010AJ....140.1868W} {140, 1868}

\bibitem[\protect\citeauthoryear{{Xilouris}, {Madden}, {Galliano}, {Vigroux}
  \& {Sauvage}}{{Xilouris} et~al.}{2004}]{Xilouris04}
{Xilouris} E.~M.,  {Madden} S.~C.,  {Galliano} F.,  {Vigroux} L.,   {Sauvage}
  M.,  2004, \mn@doi [\aap] {10.1051/0004-6361:20034020}, \href
  {http://adsabs.harvard.edu/abs/2004A%26A...416...41X} {416, 41}

\bibitem[\protect\citeauthoryear{{Yi} et~al.,}{{Yi} et~al.}{2005}]{Yi05}
{Yi} S.~K.,  et~al., 2005, \mn@doi [\apjl] {10.1086/422811}, \href
  {http://adsabs.harvard.edu/abs/2005ApJ...619L.111Y} {619, L111}

\bibitem[\protect\citeauthoryear{{Young} et~al.,}{{Young}
  et~al.}{2011}]{Young11}
{Young} L.~M.,  et~al., 2011, \mn@doi [\mnras]
  {10.1111/j.1365-2966.2011.18561.x}, \href
  {http://adsabs.harvard.edu/abs/2011MNRAS.414..940Y} {414, 940}

\bibitem[\protect\citeauthoryear{{Zibetti}, {Charlot}  \& {Rix}}{{Zibetti}
  et~al.}{2009}]{Zibetti09}
{Zibetti} S.,  {Charlot} S.,   {Rix} H.-W.,  2009, \mn@doi [\mnras]
  {10.1111/j.1365-2966.2009.15528.x}, \href
  {http://adsabs.harvard.edu/abs/2009MNRAS.400.1181Z} {400, 1181}

\bibitem[\protect\citeauthoryear{{di Serego Alighieri} et~al.,}{{di Serego
  Alighieri} et~al.}{2013}]{diSeregoAlighieri13}
{di Serego Alighieri} S.,  et~al., 2013, \mn@doi [\aap]
  {10.1051/0004-6361/201220551}, \href
  {http://adsabs.harvard.edu/abs/2013A%26A...552A...8D} {552, A8}

\bibitem[\protect\citeauthoryear{{van Dokkum} \& {Conroy}}{{van Dokkum} \&
  {Conroy}}{2012}]{vanDokkum12}
{van Dokkum} P.~G.,  {Conroy} C.,  2012, \mn@doi [\apj]
  {10.1088/0004-637X/760/1/70}, \href
  {http://adsabs.harvard.edu/abs/2012ApJ...760...70V} {760, 70}

\bibitem[\protect\citeauthoryear{{van Loon}, {Groenewegen}, {de Koter},
  {Trams}, {Waters}, {Zijlstra}, {Whitelock}  \& {Loup}}{{van Loon}
  et~al.}{1999}]{vanLoon99}
{van Loon} J.~T.,  {Groenewegen} M.~A.~T.,  {de Koter} A.,  {Trams} N.~R.,
  {Waters} L.~B.~F.~M.,  {Zijlstra} A.~A.,  {Whitelock} P.~A.,   {Loup} C.,
  1999, \aap, \href {http://adsabs.harvard.edu/abs/1999A%26A...351..559V} {351,
  559}

\makeatother
\end{thebibliography}
